\documentclass[11pt]{article}
\usepackage{PRIMEarxiv}
\usepackage[utf8]{inputenc}
\usepackage[T1]{fontenc}
\usepackage{mathptmx}
\usepackage[scaled=0.92]{helvet}
\usepackage{courier}
\usepackage{microtype}
\usepackage{amsmath,amssymb}
\usepackage{booktabs}
\usepackage{multirow}
\usepackage{array}
\usepackage{graphicx}
\usepackage{xurl}
\usepackage{xcolor}
\usepackage[inline]{enumitem}
\usepackage{caption}
\usepackage{subcaption}
\usepackage[numbers,sort&compress]{natbib}
\usepackage{tikz}
\usetikzlibrary{arrows.meta,positioning,shapes.geometric,calc}
\usepackage{hyperref}
\usepackage[nameinlink,capitalise]{cleveref}

\hypersetup{pdftitle={Integrating Slides, Mathematics, and Simulation in One Page: GenAI-Assisted HTML Learning Objects for Engineering Courses},pdfauthor={Xuan-Quy Dao et al.},colorlinks=true,linkcolor=blue!60!black,citecolor=blue!60!black,urlcolor=blue!60!black}

\newcommand{\doi}[1]{\href{https://doi.org/#1}{doi:\,\nolinkurl{#1}}}
\newcolumntype{L}[1]{>{\raggedright\arraybackslash}p{#1}}
\newcolumntype{C}[1]{>{\centering\arraybackslash}p{#1}}
\setlist{itemsep=2pt,topsep=3pt}

\newif\ifmatlabcolleague
\matlabcolleaguetrue
\newcommand{\MATLABrelease}{R2024a}
\newcommand{\MATLABmaxdiff}{0.22}

\title{Integrating Slides, Mathematics, and Simulation in One Page:\\ GenAI-Assisted HTML Learning Objects\\ for Engineering Courses}

\author{
	Xuan-Quy~Dao\textsuperscript{1}\thanks{Corresponding author:
		\texttt{quy.dao@eiu.edu.vn}.},
	Ngoc-Bich~Le\textsuperscript{2},
	Xuan-Dung~Phan\textsuperscript{1},
	and
	Le-Minh~Nguyen\textsuperscript{1}\\[4pt]
	{\small \textsuperscript{1}Eastern International University, Ho Chi Minh City, Vietnam}\\
	{\small \textsuperscript{2}Industrial University of Ho Chi Minh City, Ho Chi Minh City, Vietnam}\\[4pt]
	{\small \texttt{quy.dao@eiu.edu.vn}; \texttt{lengocbich@iuh.edu.vn};}\\
	{\small \texttt{dung.phan@eiu.edu.vn}; \texttt{minh.nguyenle@eiu.edu.vn}}
}

\begin{document}
\maketitle

\begin{abstract}
Engineering courses spread their materials across slides (PPTX), typeset notes (LaTeX/PDF), and MATLAB, leaving students to connect them. We describe an instructor-in-the-loop workflow in which generative AI (GenAI) converts these assets into single-file HTML learning objects that combine presentation, mathematics, visualization, simulation, and assessment; a page is released only after its simulations pass numerical verification. Six open demonstrators implement 18 tasks from six engineering courses. In the eight tasks with quantitative cross-checks, every quantity met its tolerance at default settings (1\%, or 2\% for the finite-difference time-domain simulation) and agreed with a separate Python re-computation and a value-blinded MATLAB re-computation run by a colleague not involved in development; the checks also flagged under-resolved settings. In a perception survey, students (19 after post-hoc screening of 28 responses) reported that the pages linked equations with simulation, and 74\% preferred them to separate files, but usability was marginal (System Usability Scale 56.8). Twelve lecturers rated the approach as useful (item means 4.25--4.58) but were hesitant to adopt it; their median ratings on a 12-criterion capability rubric agreed moderately with the first author's ($\kappa_w=0.60$; individual median 0.31). The results show technical feasibility and preliminary positive perceptions; learning outcomes were not measured.
\end{abstract}


\section{Introduction}
\label{sec:intro}

Engineering students must connect mathematical models, physical concepts, numerical methods, and visual intuition, yet these usually arrive through separate channels: concepts on PowerPoint slides (PPTX), derivations in LaTeX/PDF, and numerical work in MATLAB \citep{ulaby2015,lonngren2007,rumpf2022}. Each tool does its own job well; the difficulty lies in the gaps between them. When mutually referring sources are physically separated, learners must hold one in working memory while searching the other, the \emph{split-attention} effect of cognitive-load theory \citep{sweller1988,chandler1992}. Placing related words, equations, and pictures together reduces this load \citep{mayer2003nine,mayer2009}, and multiple representations help only when learners are supported in relating them \citep{ainsworth2006}.

A single HTML page can host all these components \citep{whatwgHTML}: typeset equations \citep{cervone2012}, 2D and 3D graphics (SVG, Canvas, or WebGL libraries such as Three.js \citep{threejs}), JavaScript simulations, and self-assessment, with no installation. Well-designed simulations support conceptual learning \citep{wieman2008,dejong1998,rutten2012}, and interactive engagement improves STEM outcomes \citep{hake1998,freeman2014}. Building such pages, however, requires web-programming skills that few engineering lecturers use routinely. Large language models now generate code, mathematics, and text \citep{chen2021codex,becker2023,prather2023robots}, and their use for teaching materials is being explored \citep{kasneci2023,lo2023,qadir2023,johri2023}. We use GenAI as a development assistant whose output is checked before release.

\paragraph{Scope of the claim.} Learning outcomes depend mainly on the instructional method, not on the medium \citep{clark1983}. We therefore do \emph{not} claim that HTML is better than PPTX, LaTeX, or MATLAB: MATLAB and Python provide validated numerical libraries that browser JavaScript does not match, and LaTeX remains the standard for typeset mathematics. Our proposition is narrower: \emph{HTML can provide an integrated learning layer that combines presentation, mathematics, visualization, simulation, and assessment.} We support it with working demonstrators, numerical verification, a capability analysis, and a perception survey. Learning effectiveness is not tested.

The study addresses three research questions:
\begin{itemize}[leftmargin=3.2em]
	\item[\textbf{RQ1:}] Can an instructor-in-the-loop GenAI workflow turn existing PPTX, LaTeX, and MATLAB materials into integrated HTML learning objects whose simulations reproduce analytical and reference solutions within explicit tolerances?
	\item[\textbf{RQ2:}] What capability profile do common teaching-material formats show across 12 anchored criteria, and where do they complement one another?
	\item[\textbf{RQ3:}] How do students and lecturers perceive the usability, learning support, adoption effort, and trustworthiness of these learning objects?
\end{itemize}

Systems that generate interactive educational content with GenAI already exist (\cref{sec:related}). What has received little attention is a process for converting an instructor's own assets into such content and checking it numerically against independent references before release. The contributions are:
\begin{enumerate}[leftmargin=1.8em]
	\item An instructor-in-the-loop workflow that uses this protocol as its release gate (\cref{fig:workflow}), with 18 tasks and six open, single-file demonstrators (\cref{sec:framework}).
	\item A 12-criterion anchored rubric that profiles five formats without ranking them (\cref{sec:rubric}).
	\item A pre-release verification protocol for GenAI-generated teaching simulations, with explicit criteria per type of check, verification visible to students, and cross-checks against analytical references, a Python re-computation, and a value-blinded MATLAB re-computation (\cref{sec:verification}).
	\item Student and lecturer perception results from 28 students (19 after screening) and 12 lecturers (\cref{sec:survey,sec:results}).
\end{enumerate}

\section{Background and Related Work}
\label{sec:related}

\paragraph{Conventional materials and notebooks.} Slides support sequential explanation but give students little chance to change a parameter; LaTeX renders mathematics precisely but produces a static PDF; MATLAB is widely used for computation and simulation \citep{lonngren2007,rumpf2022}. Integration of text, mathematics, code, and output is not new. MATLAB Live Scripts \citep{mathworksLive} have improved the student experience when paired with automated grading \citep{ni2022}, and Jupyter notebooks are a common format for computational teaching \citep{kluyver2016,barba2019}. Both differ from a single-file HTML page mainly in their dependence on a runtime (a licensed MATLAB installation or a Python kernel) and in interaction that typically requires re-executing a cell. JupyterLite narrows this gap by running kernels in the browser \citep{jupyterlite}; we evaluate Jupyter in its common hosted-kernel configuration (\cref{sec:rubric}). Interactive web-native textbooks have been associated with higher learning gains than static online text \citep{edgcomb2014}.

\paragraph{Interactive simulations.} Research-based simulations such as PhET deepen conceptual understanding \citep{wieman2008}; simulations generally enhance traditional instruction, especially within a structured, scaffolded activity \citep{rutten2012,dejong1998}. In electromagnetics, animated packages have long visualized fields and waves \citep{leung2001}. These tools are usually built by dedicated teams and sit outside a course's own notes and assessments. Our focus is on placing simulation next to the course's own text, mathematics, and questions.

\paragraph{GenAI in engineering and computing education.} GenAI can produce explanations, mathematics, code, exercises, and feedback \citep{kasneci2023,lo2023}, and may help lecturers create materials \citep{qadir2023,johri2023}. Code-generating models solve typical programming exercises \citep{chen2021codex,becker2023}, prompting a rethink of curricula \citep{denny2024,prather2023robots}, and AI pair programmers shorten task completion for professional developers \citep{peng2023copilot}. Closest to our work, MAIC-UI generates interactive STEM courseware from textbooks, slides, and PDFs through a no-code interface \citep{tu2026maic}, and Google Research's ``learning interactives'' generate scaffolded STEM simulations through a multi-stage pipeline with automated critique of the code \citep{kovshov2026interactives}. Both are recent preprints that keep the educator in the loop. The present study differs in its inputs (an engineering course's own PPTX, LaTeX, and MATLAB assets), in making numerical verification against independent references a condition of release, and in its evaluation across six engineering courses. Verification matters because language models can produce fluent but unfounded output \citep{ji2023hallucination}, and code that passes a model's own tests can fail under more rigorous testing \citep{liu2023evalplus}.

\paragraph{Verification and survey instruments.}\label{sec:vv} The computational-science literature distinguishes \emph{code verification} (does the program solve its equations correctly?), \emph{solution verification} (is a particular run numerically resolved?), and \emph{validation} (do the equations represent physical reality?) \citep{oberkampf2010}. Our checks fall under the first two; textbook models and analytical solutions are treated as reference problems, and physical validation is out of scope. For the survey we use the System Usability Scale (SUS), a ten-item questionnaire scored 0--100 with published benchmarks \citep{brooke1996,bangor2008}, and the Technology Acceptance Model (TAM), which explains adoption through perceived usefulness and ease of use \citep{davis1989,granic2019}. Both rely on Likert-type items \citep{likert1932}, which can be summarized parametrically when aggregated into scales \citep{norman2010}.

\section{Workflow, Tasks, and Demonstrators}
\label{sec:framework}

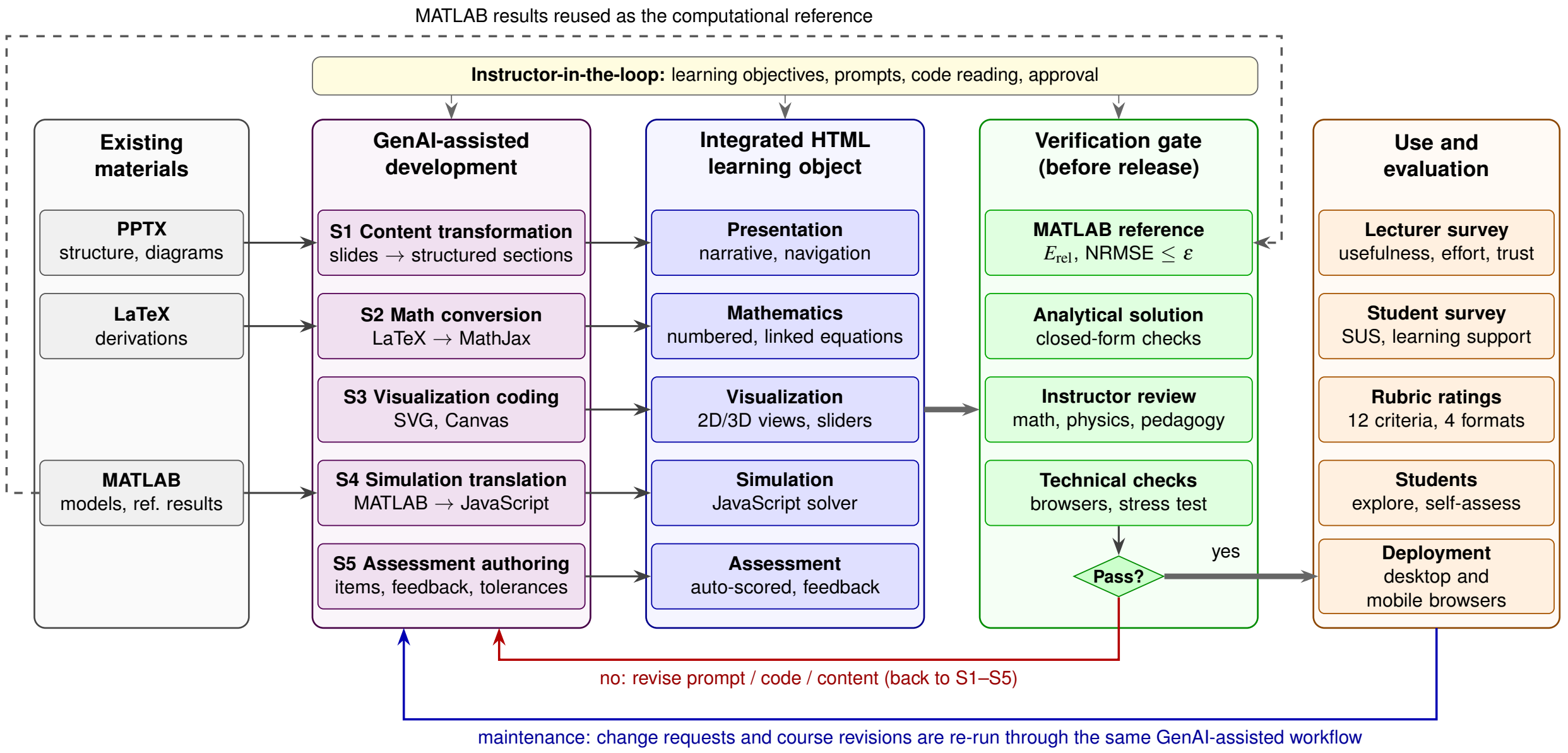
\begin{figure}[t]
	\centering
	\resizebox{\textwidth}{!}{%
		\begin{tikzpicture}[
  >={Stealth[length=2.2mm,width=1.6mm]},
  every node/.style={font=\sffamily\fontsize{7.3}{8.4}\selectfont, execute at begin node={\hyphenpenalty=10000\exhyphenpenalty=10000}},
  box/.style={draw=#1!65!black, fill=#1!12, rounded corners=2pt, align=center, inner xsep=3pt, inner ysep=2.5pt,
              minimum height=0.82cm},
  cont/.style={draw=#1!55!black, fill=#1!4, rounded corners=4pt, line width=0.6pt},
  ttl/.style={font=\sffamily\bfseries\fontsize{8.2}{9.4}\selectfont, align=center, text width=3.3cm},
  flow/.style={->, line width=0.7pt, draw=black!75},
  bigflow/.style={->, line width=2pt, draw=black!60},
  loop/.style={->, line width=0.8pt, draw=#1!70!black},
  note/.style={font=\sffamily\fontsize{7.3}{8.4}\selectfont, align=center}]
\draw[cont=gray] (0.20,0.15) rectangle (2.90,6.55);
\node[ttl] at (1.55,6.10) {Existing\\materials};
\draw[cont=violet] (3.70,0.15) rectangle (7.20,6.55);
\node[ttl] at (5.45,6.10) {GenAI-assisted\\development};
\draw[cont=blue] (7.90,0.15) rectangle (11.40,6.55);
\node[ttl] at (9.65,6.10) {Integrated HTML\\learning object};
\draw[cont=green] (12.10,0.15) rectangle (15.60,6.55);
\node[ttl] at (13.85,6.10) {Verification gate\\(before release)};
\draw[cont=orange] (16.30,0.15) rectangle (19.40,6.55);
\node[ttl] at (17.85,6.10) {Use and\\evaluation};
\node[box=gray, text width=2.35cm] (PPTX) at (1.55,5.00) {\textbf{PPTX}\\structure, diagrams};
\node[box=gray, text width=2.35cm] (TEX) at (1.55,3.95) {\textbf{LaTeX}\\derivations};
\node[box=gray, text width=2.35cm] (MAT) at (1.55,1.85) {\textbf{MATLAB}\\models, ref.\ results};
\node[box=violet, text width=3.15cm] (S1) at (5.45,5.00) {\textbf{S1 Content transformation}\\slides $\rightarrow$ structured sections};
\node[box=blue, text width=3.15cm] (H1) at (9.65,5.00) {\textbf{Presentation}\\narrative, navigation};
\node[box=orange, text width=2.75cm] (U1) at (17.85,5.00) {\textbf{Lecturer survey}\\usefulness, effort, trust};
\node[box=violet, text width=3.15cm] (S2) at (5.45,3.95) {\textbf{S2 Math conversion}\\LaTeX $\rightarrow$ MathJax};
\node[box=blue, text width=3.15cm] (H2) at (9.65,3.95) {\textbf{Mathematics}\\numbered, linked equations};
\node[box=orange, text width=2.75cm] (U2) at (17.85,3.95) {\textbf{Student survey}\\SUS, learning support};
\node[box=violet, text width=3.15cm] (S3) at (5.45,2.90) {\textbf{S3 Visualization coding}\\SVG, Canvas};
\node[box=blue, text width=3.15cm] (H3) at (9.65,2.90) {\textbf{Visualization}\\2D/3D views, sliders};
\node[box=orange, text width=2.75cm] (U3) at (17.85,2.90) {\textbf{Rubric ratings}\\12 criteria, 4 formats};
\node[box=violet, text width=3.15cm] (S4) at (5.45,1.85) {\textbf{S4 Simulation translation}\\MATLAB $\rightarrow$ JavaScript};
\node[box=blue, text width=3.15cm] (H4) at (9.65,1.85) {\textbf{Simulation}\\JavaScript solver};
\node[box=orange, text width=2.75cm] (U4) at (17.85,1.85) {\textbf{Students}\\explore, self-assess};
\node[box=violet, text width=3.15cm] (S5) at (5.45,0.80) {\textbf{S5 Assessment authoring}\\items, feedback, tolerances};
\node[box=blue, text width=3.15cm] (H5) at (9.65,0.80) {\textbf{Assessment}\\auto-scored, feedback};
\node[box=orange, text width=2.75cm] (U5) at (17.85,0.80) {\textbf{Deployment}\\desktop and mobile browsers};
\node[box=green, text width=3.15cm] (V1) at (13.85,5.00) {\textbf{MATLAB reference}\\$E_{\mathrm{rel}}$, NRMSE $\le\varepsilon$};
\node[box=green, text width=3.15cm] (V2) at (13.85,3.95) {\textbf{Analytical solution}\\closed-form checks};
\node[box=green, text width=3.15cm] (V3) at (13.85,2.90) {\textbf{Instructor review}\\math, physics, pedagogy};
\node[box=green, text width=3.15cm] (V4) at (13.85,1.85) {\textbf{Technical checks}\\browsers, stress test};
\node[diamond, aspect=2.2, draw=green!55!black, fill=green!20, inner sep=0.6pt, font=\sffamily\bfseries\fontsize{6.8}{8}\selectfont] (P) at (13.85,0.80) {Pass?};
\node[draw=black!60, fill=yellow!15, rounded corners=3pt, minimum width=11.90cm, minimum height=0.48cm, font=\sffamily\fontsize{7.2}{8.6}\selectfont] (INS) at (9.65,7.10) {\textbf{Instructor-in-the-loop:} learning objectives, prompts, code reading, approval};
\draw[->, dashed, draw=black!60] (5.45,6.86) -- (5.45,6.55);
\draw[->, dashed, draw=black!60] (9.65,6.86) -- (9.65,6.55);
\draw[->, dashed, draw=black!60] (13.85,6.86) -- (13.85,6.55);
\draw[flow] (PPTX.east) -- (S1.west);
\draw[flow] (TEX.east) -- (S2.west);
\draw[flow] (MAT.east) -- (S4.west);
\draw[flow] (S1.east) -- (H1.west);
\draw[flow] (S2.east) -- (H2.west);
\draw[flow] (S3.east) -- (H3.west);
\draw[flow] (S4.east) -- (H4.west);
\draw[flow] (S5.east) -- (H5.west);
\draw[bigflow] (11.40,2.90) -- (12.10,2.90);
\draw[flow] (V4.south) -- (P.north);
\draw[bigflow] (P.east) -- node[note, above=1pt, pos=0.4] {yes} (U5.west);
\draw[loop=gray, dashed] (MAT.west) -- (-0.15,1.85) -- (-0.15,7.60) -- node[note, above, pos=0.5] {MATLAB results reused as the computational reference} (15.90,7.60) -- (15.90,5.00) -- (V1.east);
\draw[loop=red] (P.south) -- (13.85,-0.25) -- node[note, below, text=red!60!black] {no: revise prompt / code / content (back to S1--S5)} (6.05,-0.25) -- (6.05,0.15);
\draw[loop=blue] (17.85,0.15) -- (17.85,-1.00) -- node[note, below, text=blue!60!black] {maintenance: change requests and course revisions are re-run through the same GenAI-assisted workflow} (4.85,-1.00) -- (4.85,0.15);
\end{tikzpicture}}
	\caption{Proposed workflow. GenAI stages S1--S5 convert existing PPTX, LaTeX, and MATLAB materials into the five layers of an integrated HTML learning object, which is released only after verification, instructor review, and technical checks; any failure returns to S1--S5. MATLAB is both the source of the model and a computational reference. The separate Python re-computation (\cref{sec:verification}) is not shown, and ``4 formats'' refers to the formats rated by lecturers.}
	\label{fig:workflow}
\end{figure}

\subsection{Workflow}

In the framework (\cref{fig:workflow}) each conventional resource keeps a role: PPTX supplies narrative and diagrams, LaTeX the mathematics, and MATLAB the computational model and reference results. HTML is the delivery layer. GenAI is used in five stages: \textbf{S1} content transformation (slides and notes become structured sections with objectives and navigation); \textbf{S2} mathematical conversion (LaTeX becomes MathJax with numbered, cross-referenced equations); \textbf{S3} visualization coding (SVG, Canvas, or WebGL components and their controls); \textbf{S4} simulation translation (MATLAB models become JavaScript, e.g., ordinary differential equation (ODE) integrators \citep{butcher2016}, the fast Fourier transform (FFT) \citep{cooley1965}, or finite-difference time-domain (FDTD) updates \citep{yee1966,taflove2005}); and \textbf{S5} assessment authoring (multiple-choice, numerical, and simulation-based items with feedback). The assembled page is then verified (\cref{sec:verification}); any failure returns to the relevant stage. Every page follows the same order (objectives, theory, mathematics, visualization, simulation, application, assessment), with equations placed beside the visualization they govern \citep{mayer2003nine} and every read-out computed from a displayed equation.

\subsection{GenAI implementation and reproducibility}
\label{sec:genai-impl}

The demonstrators were generated with a commercially available large language model. The provider, model version, configuration, and complete session logs were not retained. Reproducibility therefore has three levels: (1) \emph{running the demonstrators}, which anyone can do by opening the released files; (2) \emph{reproducing the verification and survey analysis}, which requires scripts and data that are prepared but not yet publicly archived; and (3) \emph{reproducing the GenAI authoring}, which is not possible. The empirical claims rest on levels 1 and 2.

For each course GenAI produced one HTML file: page structure and text, MathJax markup, visualization code, JavaScript solvers with verification metrics, and quiz items. The task specifications, textbook models, and tolerances were set by the instructor. Each request followed a fixed six-part structure (Appendix~\ref{app:prompts}), given as a template because the verbatim prompts were not retained. No output was accepted automatically: the instructor read the code, reviewed the mathematics and physics against the textbook, ran the verification panels at default, extreme, and random settings, and checked the pages in desktop and mobile browsers. Failures were handled by prompt revision or by manual correction. How often each occurred, and how long it took, was not recorded, so we make no claim about development effort.

\subsection{Tasks and demonstrators}
\label{sec:tasks}

Six undergraduate courses define the scope: Engineering Mathematics \citep{kreyszig2011}, Circuit Analysis \citep{nilsson2019}, Electromagnetics \citep{ulaby2015}, Signals and Systems \citep{oppenheim1997}, Control Systems \citep{nise2019,astrom2006}, and Digital Signal Processing \citep{oppenheim2010}. Three complexity levels are defined by the layers a learning object must contain:
\begin{itemize}[leftmargin=1.5em]
	\item \textbf{Easy (E):} presentation, mathematics, static diagrams, and at least three auto-scored items.
	\item \textbf{Medium (M):} E plus at least two learner-controlled parameters driving a visualization computed from closed-form expressions.
	\item \textbf{Complex (C):} M plus a numerical method (time stepping, FFT, filter design, or FDTD) and quantitative verification against a reference.
\end{itemize}
Crossing courses and levels gives the 18 tasks in \cref{tab:tasks}.

\begin{table}[!t]
	\centering
	\scriptsize
	\setlength{\tabcolsep}{3.5pt}
	\caption{Task framework (6 courses $\times$ 3 levels). Ref.: verification reference (A = analytical; M = MATLAB, i.e., the reference script embedded in the page and the value-blinded re-computation of \cref{tab:validation}). Demo: demonstrator file implementing the task. Quantitative cross-checks against Python and MATLAB are reported in \cref{tab:validation} for the eight cases marked M.}
	\label{tab:tasks}
	\begin{tabular}{L{0.95cm}C{0.4cm}L{4.1cm}L{6.4cm}C{0.7cm}C{0.8cm}}
		\toprule
		\textbf{ID} & \textbf{Lvl} & \textbf{Learning object} & \textbf{HTML components / numerical method} & \textbf{Ref.} & \textbf{Demo} \\
		\midrule
		MAT-E & E & Dot and cross products & Definitions, MathJax, SVG geometry, worked example & A & $\checkmark$ \\
		MAT-M & M & Interactive 3D vector algebra & Rotatable 3D view; live $\mathbf{A}\cdot\mathbf{B}$, $\mathbf{A}\times\mathbf{B}$, angle, projection & A & $\checkmark$ \\
		MAT-C & C & Second-order linear ODE & RK4 solver, error plot, convergence check & A, M & $\checkmark$ \\
		\midrule
		CIR-E & E & Phasors and impedance of R, L, C & Impedance plane (SVG), worked example & A & $\checkmark$ \\
		CIR-M & M & Series RLC in AC steady state & Sliders $R,L,C,f$; phasor diagram; $|I(f)|$, resonance, KVL check & A & $\checkmark$ \\
		CIR-C & C & RLC step transient & RK4 state equations; over-, critically and under-damped cases & A, M & $\checkmark$ \\
		\midrule
		EMF-E & E & Coulomb's law, field of a point charge & Vector-form equations, SVG field diagram, worked example & A & $\checkmark$ \\
		EMF-M & M & Field lines and equipotentials of $N$ point charges & Potential map, traced field lines, draggable charges and probe; $\mathbf{E}$ vs.\ $-\nabla V$ & A, M & $\checkmark$ \\
		EMF-C & C & 1D FDTD pulse at a dielectric interface & Yee update \citep{yee1966}, Mur boundaries, animation, probes; $\Gamma,\tau$ & A, M & $\checkmark$ \\
		\midrule
		SIG-E & E & Elementary signals, time operations & Shift, scaling, and reversal plots & A & $\checkmark$ \\
		SIG-M & M & Graphical convolution & Flip-and-slide animation, three signal pairs & A & $\checkmark$ \\
		SIG-C & C & Sampling, aliasing, reconstruction & DFT with Hann window, sinc interpolation, alias prediction & A, M & $\checkmark$ \\
		\midrule
		CTL-E & E & Transfer functions, poles, stability & $s$-plane diagram, worked example & A & $\checkmark$ \\
		CTL-M & M & Second-order step response & Sliders $\zeta,\omega_n$; pole map; overshoot, $T_p$, $T_s$ & A & $\checkmark$ \\
		CTL-C & C & PID speed control of a DC motor & Closed-loop RK4, partial-fraction reference, Bode plot, margins \citep{astrom2006} & A, M & $\checkmark$ \\
		\midrule
		DSP-E & E & DFT definition, frequency resolution & 8-point DFT example (stem plots) & A & $\checkmark$ \\
		DSP-M & M & Windowing and spectral leakage & JavaScript FFT, four windows, weak-tone visibility & A, M & $\checkmark$ \\
		DSP-C & C & Windowed-sinc FIR design and filtering & Low/high-pass design, filtering, gain verification & A, M & $\checkmark$ \\
		\bottomrule
	\end{tabular}
\end{table}

\label{sec:demos}For each course, one single-file HTML page implements the E, M, and C tasks and ends with a self-assessment quiz (file statistics in \cref{tab:demos}; screenshots in Appendix~\ref{app:screens}). The files share a layer of about 200 lines: a plotting class, a classical fourth-order Runge--Kutta (RK4) integrator \citep{butcher2016}, a radix-2 FFT \citep{cooley1965}, the error metrics of \cref{sec:verification}, and a quiz engine. The only external dependency is MathJax 3.2.2 \citep{cervone2012}, loaded from a content delivery network, so a page needs a connection on first use. There is no framework, build step, or server, and each file is 33--41\,KB.

Three design choices recur. First, every number is shown next to the formula that produces it. Second, verification is visible: each Complex section has a panel with a PASS/FAIL verdict and a collapsible MATLAB script, and when a student coarsens the RK4 step or narrows the FDTD pulse until the error exceeds the tolerance, the verdict turns red and names the cause. (The panel was labelled ``validation'' in the survey version and ``verification'' in the released files.) Third, approximations are labelled: in CTL-M, the settling-time rule $4/\zeta\omega_n$ is compared with the measured value, and the page explains the difference.

\section{Capability Rubric}
\label{sec:rubric}

Instead of ranking formats, we rate each on 12 criteria with five behaviourally anchored levels, so that raters judge from observable features \citep{jonsson2007}. The anchors and full scoring rules are in Appendix~\ref{app:rubric}. In brief: each format is rated on its native capability as typically used in teaching, not on what our demonstrators contain; C9 concerns licences, runtimes, and devices, not accessibility for users with disabilities \citep{wcag22}; C10 counts how many of the five layers (C1, C2, C3, C6, C8) score at least 3; and no total is computed, because the criteria are not commensurable.

The first author rated five formats. Twelve lecturers (\cref{sec:survey}) independently rated four of them, with C10 rated directly. \Cref{tab:profile} shows both.

\begin{table}[!ht]
	\centering
	\small
	\setlength{\tabcolsep}{6pt}
	\caption{Capability profile on the rubric of \cref{tab:rubric} (1--5): first author's rating\,/\,lecturers' median ($N_L=12$). Jupyter was rated by the first author only. $\kappa_w$: quadratic-weighted agreement \citep{cohen1968} between the two profiles over the 12 criteria (half-point medians rounded up). C12 is reverse-coded (higher = lower authoring demand).}
	\label{tab:profile}
	\begin{tabular}{lccccc}
		\toprule
		\textbf{Criterion} & \textbf{PPTX} & \textbf{LaTeX/PDF} & \textbf{MATLAB} & \textbf{Jupyter} & \textbf{HTML} \\
		\midrule
		C1 Content organization        & 4\,/\,4   & 5\,/\,4   & 4\,/\,4   & 4 & 5\,/\,4 \\
		C2 Mathematical representation & 3\,/\,4   & 5\,/\,4   & 3\,/\,4   & 4 & 4\,/\,4 \\
		C3 Static visualization        & 4\,/\,4   & 5\,/\,4   & 4\,/\,4.5 & 4 & 4\,/\,4.5 \\
		C4 Interactive visualization   & 2\,/\,3.5 & 1\,/\,3   & 4\,/\,4   & 4 & 5\,/\,5 \\
		C5 3D visualization            & 3\,/\,3   & 2\,/\,3   & 5\,/\,4.5 & 4 & 4\,/\,4 \\
		C6 Numerical simulation        & 2\,/\,3   & 2\,/\,3   & 5\,/\,5   & 5 & 3\,/\,4 \\
		C7 Parameter interaction       & 2\,/\,3   & 1\,/\,3   & 4\,/\,5   & 4 & 5\,/\,4.5 \\
		C8 Assessment integration      & 3\,/\,3.5 & 2\,/\,3   & 3\,/\,4   & 3 & 4\,/\,4 \\
		C9 Cross-platform access       & 3\,/\,3.5 & 5\,/\,3.5 & 2\,/\,3   & 2 & 4\,/\,5 \\
		C10 Content integration        & 4\,/\,3.5 & 3\,/\,3.5 & 5\,/\,5   & 5 & 5\,/\,5 \\
		C11 Ease of updating           & 3\,/\,4   & 4\,/\,3   & 4\,/\,4   & 4 & 3\,/\,3 \\
		C12 Authoring demand (rev.)    & 5\,/\,5   & 3\,/\,3.5 & 3\,/\,3   & 3 & 2\,/\,3 \\
		\midrule
		$\kappa_w$                     & 0.46 & 0.44 & 0.72 & -- & 0.71 \\
		\bottomrule
	\end{tabular}
\end{table}

\paragraph{First author's profile.} HTML is not the strongest format everywhere: LaTeX leads on mathematics and static figures, and MATLAB and Jupyter on numerical simulation. Integration alone does not distinguish HTML either, because Live Scripts and Jupyter notebooks also reach C10\,$=5$. What distinguishes HTML is the \emph{combination} of integration (C10), real-time interaction (C4, C7), and access without a licence or conventional runtime (C9): under the configurations rated, it is the only format scoring at least 4 on all four. This depends on the scope rule for Jupyter; a browser-only distribution such as JupyterLite \citep{jupyterlite} would score higher on C9. The main weakness of HTML is authoring: updating a page requires specialist web skills (C11), and building one requires two or three languages (C12).

\paragraph{Lecturers' ratings.} Agreement between the lecturers' median profile and the first author's over all 48 cells was moderate ($\kappa_w=0.60$), highest for MATLAB and HTML and lowest for LaTeX/PDF. This is agreement of an aggregated profile, not inter-rater reliability: agreement between individual lecturers and the first author varied widely (median $\kappa_w=0.31$, range $-0.04$ to $0.74$). Two lecturers gave all four formats the same score on every criterion; excluding them changes the profile agreement only to 0.63. Rating HTML 4 on C6 gives 0.61, and omitting C10 gives 0.58.

The lecturers' profiles are flatter. They rated LaTeX/PDF lower on organization, mathematics, and static figures and higher on interactivity and simulation, and they rated HTML higher on numerical simulation (4 against 3) and cross-platform access (5 against 4). Two results bear on the claims above. First, the lecturers' medians reproduce the observation that HTML is the only rated format scoring at least 4 on C4, C7, C9, and C10; MATLAB receives 3 for C9. Second, lecturers agree that authoring is the weak side of HTML: their median on C11 is 3, and only 5 of 12 rated HTML at least as high as the other formats on C12.

\section{Verification Protocol and Results}
\label{sec:verification}

\subsection{Metrics, tolerances, and release conditions}

Each GenAI-generated simulation undergoes code verification and, for each parameter setting, solution verification \citep{oberkampf2010}. The reference is an analytical solution where one exists and otherwise a MATLAB computation. For a scalar quantity $X$ we use the relative error, and for a waveform $x[k]$ sampled at $K$ points the range-normalized root-mean-square error:
\begin{equation}
	E_{\mathrm{rel}}=\frac{|X_{\mathrm{HTML}}-X_{\mathrm{ref}}|}{|X_{\mathrm{ref}}|}\times100\%,\qquad
	\mathrm{NRMSE}=\frac{\sqrt{\frac{1}{K}\sum_{k}\left(x_{\mathrm{HTML}}[k]-x_{\mathrm{ref}}[k]\right)^2}}{\max_k x_{\mathrm{ref}}[k]-\min_k x_{\mathrm{ref}}[k]}\times100\%.
\end{equation}
Normalizing by the range keeps the metric defined when the reference crosses zero. For magnitudes displayed in decibels, relative error is taken on the linear amplitude, not on the dB number, and absolute dB differences are reported separately. \Cref{tab:criteria} in Appendix~\ref{app:verif} lists the criterion for each kind of check.

The default tolerance is $\varepsilon=1\%$, with $\varepsilon=2\%$ for FDTD to allow for the numerical dispersion of the Yee scheme \citep{taflove2005}. This is a teaching-oriented threshold, not a general criterion of accuracy: an error of 1\% of the displayed range is about three pixels on a 300-pixel axis, yet the threshold still exposes under-resolved settings. A suitable value depends on the quantity, the method, and the instructional objective, and only half of the lecturers considered 1\% appropriate (\cref{sec:results}).

A learning object is released only if
\begin{enumerate*}[label=(\alph*)]
	\item all layers required by its level are present;
	\item it runs without console errors in current desktop Chrome, Firefox, and Edge and remains usable on a mobile browser;
	\item every verification metric meets its criterion at the default settings;
	\item a stress test (each control at its extremes and 40 random parameter combinations per file) raises no runtime errors, and under-resolved settings produce a visible FAIL with an explanation; and
	\item the instructor has approved the mathematics and physics.
\end{enumerate*}
A FAIL at a deliberately coarse setting is the expected behaviour of solution verification and does not block release.

\subsection{Cross-checks and their independence}

Each reference value was checked by three routes. \emph{Analytical references} establish agreement with known solutions where these exist. A \emph{Python re-computation} (SciPy for ODEs and filters, a separate NumPy implementation of the Yee scheme) is a separate implementation, but by the developer of the pages, so it cannot exclude a shared error \citep{liu2023evalplus}. A \emph{MATLAB cross-check} adds independence of the person and, for most quantities, of the numerical method: a colleague who took no part in development executed a developer-prepared package in MATLAB \MATLABrelease{} on a separate computer, without seeing the values on the pages, using MATLAB's own solvers and toolbox functions wherever possible. Agreement was decided by a script using a rule fixed before the run (Appendix~\ref{app:verif}).

These routes are complementary, not fully independent. The developer prepared the MATLAB package, the checker did not write the reference implementations, and two quantities (the RK4 order test and the FDTD coefficients) reuse the same numerical schemes.

\subsection{Results}

\Cref{tab:verifsummary} summarizes the eight tasks with quantitative cross-checks: all six Complex tasks and two Medium tasks. The other Medium tasks carry analytical checks inside the pages only. Every case passed. The Python values agreed with the in-page values to the displayed precision, and all 31 quantities in the MATLAB comparison agreed with the in-page values (largest relative difference \MATLABmaxdiff\%, for an RK4 error ratio displayed with one decimal). Two results are code verification in the strict sense: in MAT-C, halving the step divided the RK4 error by 16.3 and then 16.2, matching the theoretical order of four; in EMF-M, the field obtained by superposition matched the numerical gradient $-\nabla V$ to within $2.5\times10^{-7}\%$. The full values are in \cref{tab:validation}.

\begin{table}[!t]
	\centering
	\small
	\caption{Verification summary at default parameters (full values and notes in \cref{tab:validation}). In-page error: JavaScript simulation against the analytical or reference solution. $\Delta$: largest difference between the value-blinded MATLAB re-computation and the displayed in-page values. All rows PASS ($\varepsilon=1\%$; 2\% for EMF-C).}
	\label{tab:verifsummary}
	\begin{tabular}{L{1.1cm}L{4.4cm}L{5.0cm}L{3.3cm}}
		\toprule
		\textbf{Task} & \textbf{Quantity} & \textbf{In-page error vs.\ reference} & \textbf{$\Delta$ (MATLAB vs.\ page)} \\
		\midrule
		MAT-C & $y(t)$ by RK4, $h=0.1$\,s; error ratio for $h\to h/2$ & NRMSE $8.0\times10^{-4}$\%; ratios 16.3, 16.2 (theory 16) & $\le1.9\times10^{-3}$\%; ratio $\le\MATLABmaxdiff$\% \\
		CIR-C & RLC step response $v_C(t)$ & NRMSE $4.1\times10^{-6}$\% & NRMSE $4.1\times10^{-6}$\% \\
		EMF-M & $\mathbf{E}$ at a probe (dipole) vs.\ $-\nabla V$ & residual $2.5\times10^{-7}$\% & $\le0.03$\% \\
		EMF-C & FDTD $\Gamma$, $\tau$ at $\varepsilon_r=4$ & $E_{\mathrm{rel}}\le0.16$\% ($\le1.02$\% for $\varepsilon_r\le12$) & $\le0.02$\% \\
		SIG-C & Alias frequency & $E_{\mathrm{rel}}=1.6\times10^{-4}$\% & $\le2.5\times10^{-4}$\% \\
		CTL-C & PID closed-loop step response & NRMSE $1.2\times10^{-7}$\% & $\le0.18$\% \\
		DSP-M & Peak sidelobe of four windows & within 0.06\,dB of large-$N$ theory & $\le0.005$\,dB \\
		DSP-C & FIR gain $|H|$ at three frequencies & equal to displayed precision & $\le0.014$\%; $\le0.002$\,dB \\
		\bottomrule
	\end{tabular}
\end{table}

\paragraph{Stress test.} The stress test raised no runtime errors. It did produce FAIL verdicts, each for an under-resolved run: for example, MAT-C with step $h=0.6$\,s (NRMSE 1.12\%), and EMF-C with a narrow pulse in a dense dielectric ($E_{\mathrm{rel}}(\tau)=27\%$ from numerical dispersion). The checks therefore distinguish a correct implementation used beyond its resolution limits from an incorrect one, and students can see the difference on the page. The random seed and sampling distribution were not recorded, so the test documents the configurations reached, not a reproducible sample, and it cannot exclude an incorrect PASS at an untested setting.

One defect was found after the survey and was not reached by the stress test: CIR-C divided the peak error by a reference peak that can lie near 0\,V, producing a spurious FAIL. The released files normalize by $\max(|v_{\mathrm{pk}}|,\ \text{range of }v_C)$; default-setting values are unchanged (Appendix~\ref{app:verif}).

\paragraph{What a PASS means.} A PASS means that a quantity met the stated numerical criterion at the tested settings. It is not experimental validation of the textbook model, and it says nothing about untested settings.

\section{Survey Method}
\label{sec:survey}

The educational evaluation is a cross-sectional perception survey. It is not a controlled experiment and measures neither learning gains nor development time.

\paragraph{Participants and procedure.} All participants were recruited at Eastern International University (EIU), the institution of three of the authors. Both questionnaires were delivered online in Vietnamese between 2 and 27 October 2026, anonymously and voluntarily, with a consent box ticked before the items were shown. We received 28 responses from undergraduate engineering students, each of whom had worked through at least one demonstrator in class or at home. Twelve lecturers of engineering courses, colleagues of three of the authors, reviewed the six demonstrators, the workflow, and the embedded MATLAB scripts, then completed a questionnaire that included the rubric rating of \cref{sec:rubric}. Demographics are in Appendix~\ref{app:itemstats}. Two points matter for interpretation: most students who reported it had opened a single demonstrator (13 of 16), often for under 30 minutes (11 of 16), and five of the twelve lecturers reported no web-programming skill.

\paragraph{Data screening (post hoc).} After inspecting the responses, we excluded student responses in which at least 25 of the 26 Likert answers were identical, a standard marker of careless responding \citep{meade2012careless}. This rule was not pre-specified. It excludes nine responses (seven at ``5'' and two at ``3''), leaving 19 for the primary analysis; a sensitivity analysis on all 28 is in \cref{tab:sensitivity}. No lecturer response met the criterion.

\paragraph{Instruments and analysis.} Both questionnaires use five-point Likert items \citep{likert1932} followed by open-ended questions (constructs and items in Appendix~\ref{app:items}). Students completed the SUS and five project scales: perceived integration (SI), learning support (SL), engagement (SE), trust in simulation (ST), and preference (SP). Lecturers completed six scales: usefulness (LU) and ease of adoption (LE), based on TAM, and development effort with GenAI (LD), maintainability (LM), trust and verification (LT), and intention to use (LI). For each construct we report the mean, standard deviation, median, percentage of agreement (scores 4--5), and Cronbach's $\alpha$ \citep{cronbach1951}, with $\alpha\ge0.70$ taken as acceptable. Differences from the midpoint (3) are tested with the one-sample Wilcoxon signed-rank test on construct means \citep{wilcoxon1945}, with Holm correction \citep{holm1979}; implementation details are in Appendix~\ref{app:itemstats}. Rubric ratings are compared using quadratic-weighted $\kappa$ \citep{cohen1968}. This plan was specified before data collection; only the screening rule was added afterwards.

\section{Survey Results}
\label{sec:results}

\begin{table}[t]
	\centering
	\small
	\setlength{\tabcolsep}{5pt}
	\caption{Survey results. Likert scale 1--5; Agree: share of item responses 4--5; $p$: one-sample Wilcoxon test of construct means against 3, Holm-adjusted within each questionnaire. LM3 is reverse-scored. For lecturer scales with $\alpha<0.70$ (LU, LD, LM, LT), the composites are shown for transparency only and are interpreted item by item (\cref{tab:items}).}
	\label{tab:survey}
	\begin{tabular}{lcccccc}
		\toprule
		\textbf{Construct} & \textbf{$\alpha$} & \textbf{Mean} & \textbf{SD} & \textbf{Median} & \textbf{Agree (\%)} & \textbf{$p$} \\
		\midrule
		\multicolumn{7}{l}{\emph{Students, quality-screened ($N_S=19$)}}\\
		SI Perceived integration          & 0.80 & 3.99 & 0.64 & 3.75 & 72 & 0.001 \\
		SL Perceived learning support     & 0.85 & 3.95 & 0.61 & 4.00 & 74 & 0.001 \\
		SE Engagement and interactivity   & 0.82 & 4.02 & 0.60 & 4.00 & 77 & 0.001 \\
		ST Trust in simulation            & 0.93 & 4.04 & 0.69 & 4.00 & 75 & 0.001 \\
		SP Preference for integrated page & 0.91 & 4.00 & 0.75 & 4.00 & 76 & 0.001 \\
		SUS score (0--100)                & 0.57 & 56.8 & 8.9  & 55.0 & \multicolumn{2}{l}{marginal \citep{bangor2008}} \\
		\midrule
		\multicolumn{7}{l}{\emph{Lecturers ($N_L=12$)}}\\
		LU Perceived usefulness                    & 0.55    & 4.44 & 0.36 & 4.50 & 98 & 0.012 \\
		LE Perceived ease of adoption              & 0.80    & 3.73 & 0.70 & 3.75 & 65 & 0.020 \\
		LD Perceived development effort with GenAI & 0.65    & 4.02 & 0.43 & 4.00 & 79 & 0.012 \\
		LM Perceived maintainability               & 0.35    & 3.25 & 0.64 & 3.17 & 47 & 0.187 \\
		LT Trust and verification                  & $-0.15$ & 3.81 & 0.36 & 3.83 & 67 & 0.012 \\
		LI Intention to use                        & 0.82    & 3.67 & 0.58 & 3.50 & 58 & 0.020 \\
		\bottomrule
	\end{tabular}
\end{table}

\Cref{tab:survey} summarizes both questionnaires; \cref{tab:items} gives every item.

\subsection{Students}

All five project constructs had acceptable internal consistency ($\alpha=0.80$--$0.93$) and means of 3.95--4.04, with 72--77\% of item responses in agreement. In a convenience sample of 19, the tests against the midpoint describe the direction of opinion in this group and do not support generalization. The items with the highest agreement concerned integration: 89\% could see how the equations related to the simulation (SI2), and 84\% agreed that one page saved them from switching between files (SI1). The weakest project item was SI4, ``I rarely lost track of where I was in the topic'' (47\% agreed). Seventy-four percent preferred the integrated page to separate files (SP1). With all 28 responses, every construct mean is about 0.2 higher and the interpretation is unchanged.

Usability was less favourable. The SUS score had a mean of 56.8 (95\% CI 52.5--61.1); 13 of the 19 scores fell in the ``marginal'' band of 50--70 \citep{bangor2008}, well below the cross-study average of about 70. Two features make this number hard to read at face value. Internal consistency was low ($\alpha=0.57$), and the mean of the five positively worded items was almost uncorrelated with that of the five negatively worded ones ($r=0.03$): students who agreed that the page was easy to use did not, on the whole, disagree that it was complex. The clearest case is SUS4, ``I would need the support of a technical person'', with which 74\% agreed. Mixed wording is a known source of misreading in SUS \citep{sauro2011positive}, and in Vietnamese SUS4's ``person with technical expertise'' can be read as a lecturer or subject expert, so wording and translation may have contributed.
A tendency to agree regardless of wording would produce the same pattern and, because all project items are positively worded, would also raise the construct means above.
The agreement with SUS4 is nevertheless consistent with SI4 and with the open-ended comments: several students felt they would need guidance to find their way through a page that holds this much material.

\subsection{Lecturers}

Four lecturer scales (LU, LD, LM, LT) fell below $\alpha=0.70$, and LT was negative, so we interpret individual items (\cref{tab:items}); LE and LI can be read as scales.

\emph{Usefulness.} Every lecturer agreed with LU1, LU2, and LU4, and 92\% with LU3 (item means 4.25--4.58).
\emph{Effort.} Most expected GenAI to reduce their own programming (LD2, 83\%) and judged that building such a page without GenAI would take too long (LD3, 92\%), but fewer agreed that building one with GenAI would take an acceptable amount of their time (LD1, 58\%). These are expectations of lecturers who reviewed finished pages; none built one.
\emph{Adoption and maintenance.} Ease of adoption was lower than usefulness (LE mean 3.73). Two-thirds agreed that updating a page with GenAI would be manageable (LM1), and two-thirds were also concerned about keeping pages working as browsers and libraries change (LM3).
\emph{Trust.} The three LT items ask different questions and are reported separately: 83\% would not release a GenAI-generated simulation without checks of this kind (LT3), 67\% said the comparison with references made them trust the generated code (LT1), and only half considered a 1\% tolerance appropriate (LT2).
\emph{Intention.} Three-quarters would recommend the approach to colleagues (LI2), but only five of twelve intended to try it in their own courses (LI1); the others were neutral.

\subsection{Open-ended responses}

Few students wrote substantive comments. Three of the 19 named visualization as the most helpful feature, and two described information overload, for example ``it needs a teacher to guide its use, because there is a lot of information and it is easy to get lost''. One screened-out respondent reported ``still many display errors''. Ten lecturers named obstacles to adoption: skills or familiarity (four), time to build, test, and update content (two), cost of licences and AI tokens (two), and the evidence base (two). One lecturer criticized the content itself: theory not systematically organized, inconsistent notation, unexplained symbols, and wording that was hard to read. Quotations are translated from Vietnamese.

\section{Discussion}
\label{sec:discussion}

\paragraph{RQ1: feasibility and correctness.} A five-stage, instructor-in-the-loop workflow produced six single-file pages implementing all 18 tasks. All quantities in the verification protocol met their criteria at default settings, the RK4 solver showed fourth-order convergence, and under-resolved runs were flagged as failures. GenAI does not raise the ceiling of what HTML can do, since a skilled web developer could write the same pages by hand; it served as a development assistant whose output had to be verified. Because the verification panels and numerical parameters are on the page, verification is also something students can explore: they can change the RK4 step, the FDTD pulse width, or the sampling rate and watch the verdict turn to FAIL. Whether this improves their understanding of numerical methods was not assessed.

\paragraph{RQ2: an integration layer, not a replacement.} No format is strongest on every criterion. HTML scores below LaTeX on typeset mathematics and below MATLAB and Jupyter on numerical simulation; what it adds, in both the first author's and the lecturers' profiles, is the combination of all five layers, real-time interaction, and access on common devices. Each tool keeps a role: PPTX for instructor-led lectures, LaTeX for printed notes and examinations, and MATLAB for advanced computation and as the computational reference. Following \citet{clark1983}, any learning benefit must come from the instructional design the format makes possible, such as contiguity of equation and simulation \citep{mayer2003nine} and scaffolded exploration \citep{dejong1998}, not from the format itself.

\paragraph{RQ3: perceptions.} Students agreed most strongly with the items that describe split attention being removed, yet gave the pages a marginal usability score. We do not think these findings conflict. Putting five layers and three levels on one page removes the search between files but creates a long page in which a student can lose their place: split attention across files becomes a navigation load within the page. Scaffolding matters for simulation-based learning \citep{dejong1998,rutten2012}, and the pages were mostly used without it. The next version should add a formula index, clearer within-page navigation, and a short guided activity for each Complex section.

Lecturers rated the approach as useful but were less ready to adopt it; the obstacles they named concerned skills, time, money, and evidence. They endorsed the verification rule more than its threshold, so the tolerance should be set and justified per quantity and per course. One lecturer's criticism of inconsistent notation points to a weakness the protocol does not cover: the checks confirm that the numbers are right, not that the explanatory text is consistent and readable. Instructor review has to include that.

\paragraph{What is and is not demonstrated.} The evidence supports technical feasibility across six courses, numerical correctness at the tested settings, and positive perceptions in this sample, with the reservations above. It does not show improved learning, reduced cognitive load, shorter development time, adoption at scale, or long-term maintainability. Students' sense of how much they learned can diverge from what they actually learned \citep{deslauriers2019}, so positive perceptions are a precondition for adoption, not evidence of effectiveness. All six courses rest on well-defined analytical models; laboratory-intensive, hardware-oriented, or data-driven courses would need the workflow and its verification to be adapted and evaluated separately.

\section{Limitations}
\label{sec:limitations}

\begin{enumerate}[leftmargin=1.8em]
	\item \textbf{No controlled experiment.} There were no comparison conditions and no pre- or post-tests, so learning gains \citep{hake1998} remain unmeasured.
	\item \textbf{Survey design.} Both samples are small, self-selected, and from a single institution. The lecturers are colleagues of three authors, and Electromagnetics, the largest student group, is taught by the first author, so social-desirability bias is possible; descriptively, the six Electromagnetics students gave lower, not higher, ratings than the ten others on four of the five scales (differences of $-0.35$ to $-0.49$) and a lower SUS score (54.6 against 58.3). Exposure was brief, and novelty may inflate ratings. The screening rule was post hoc, although conclusions hold with or without it. The SUS was used in translation and behaved inconsistently. All project items are positively worded, so acquiescence cannot be separated from genuine agreement. Four lecturer scales did not reach $\alpha=0.70$, and the questionnaires need revision and piloting.
	\item \textbf{Rubric.} The first author's ratings come from a rater with an interest in the HTML approach. Criteria partly overlap, and scores depend on the stated scope. Accessibility in the sense of WCAG \citep{wcag22} was not assessed. The lecturer form described HTML as a page ``like the demonstrators'', so lecturers may have rated the demonstrators, not the format.
	\item \textbf{Independence of verification.} The three routes are complementary, not fully independent (\cref{sec:verification}). The MATLAB cross-check was performed by one colleague on one release using a developer-prepared package. Only eight of the 18 tasks were cross-checked, at default settings.
	\item \textbf{GenAI reproducibility.} The model, configuration, session logs, and manual interventions were not recorded. The paper therefore gives no documented example of a generation error and its correction, and development effort was not measured.
	\item \textbf{Technical scope and data sharing.} Browser simulation does not replace MATLAB or Python for intensive computation, and the pages need maintenance as browsers and libraries evolve. Students used repository versions up to commit a9c3c27; the released files also correct the CIR-C metric. The verification packages, analysis script, and survey responses are not yet in a permanent public archive.
\end{enumerate}

\section{Conclusion}
\label{sec:conclusion}

Within six courses and 18 tasks, this study establishes the technical feasibility of a verification-gated, GenAI-assisted workflow for producing integrated HTML learning objects from PPTX, LaTeX, and MATLAB materials. The demonstrators pass code and solution verification against analytical references, a Python re-computation, and a value-blinded MATLAB cross-check; these checks are complementary, not fully independent, and a PASS is not physical validation. The rubric profiles HTML as distinctive in combining integration, real-time interaction, and access without a conventional runtime, not as superior on every criterion. Students perceived the pages as connecting equations with simulation but rated usability as marginal and asked for guidance; lecturers found the approach useful but were hesitant to adopt it. Learning outcomes were not measured. Future work should add scaffolded navigation, revise and pilot the questionnaires, run controlled experiments with pre- and post-tests on larger samples, measure development time against manual authoring, broaden independent verification, and evaluate maintenance over time.

\section*{Data and Code Availability}
The six demonstrators and an index page are available in the public project repository at \url{https://github.com/Xdao85/GenAI_HTML}. The repository contains the English and Vietnamese demonstrator sources, an index page, questionnaire pages, collection scripts, and a README with local-use and deployment instructions. It supports level-1 reproduction as defined in \cref{sec:genai-impl}: every displayed value and verdict can be regenerated by opening the released files. The Python and MATLAB verification packages with their outputs (including the re-computation of the CIR-C grid peak in \cref{tab:validation}), the survey-analysis script, and the anonymized survey responses have been prepared but are not yet in a permanent public archive; release of the survey responses depends on confirmation of the applicable consent and data-sharing permissions.

\section*{Ethics Statement}
 Participation was anonymous and voluntary, consent was recorded before the items were shown, and no identifying information was collected (\cref{sec:survey}).




\section*{Acknowledgements}
We thank V\~o Th\'{\^e} Duy for performing the value-blinded MATLAB cross-check reported in \cref{tab:validation}.

\section*{Use of Generative AI}
As described in the paper, generative AI assisted the transformation of educational materials and the generation of HTML, JavaScript, and assessment content; the authors reviewed and checked the generated material. 

\bibliographystyle{unsrtnat}
\bibliography{references}

@book{ulaby2015,
  author    = {Ulaby, Fawwaz T. and Ravaioli, Umberto},
  title     = {Fundamentals of Applied Electromagnetics},
  edition   = {7th},
  publisher = {Pearson},
  address   = {Boston, MA},
  year      = {2015},
  isbn      = {978-0-13-335681-6}
}

@book{lonngren2007,
  author    = {Lonngren, Karl E. and Savov, Sava V. and Jost, Randy J.},
  title     = {Fundamentals of Electromagnetics with {MATLAB}},
  edition   = {2nd},
  publisher = {SciTech Publishing (IET)},
  address   = {Raleigh, NC},
  year      = {2007},
  doi       = {10.1049/SBEW043E}
}

@book{rumpf2022,
  author    = {Rumpf, Raymond C.},
  title     = {Electromagnetic and Photonic Simulation for the Beginner: Finite-Difference Frequency-Domain in {MATLAB}},
  publisher = {Artech House},
  address   = {Norwood, MA},
  year      = {2022},
  isbn      = {978-1-63081-926-2}
}

@book{kreyszig2011,
  author    = {Kreyszig, Erwin},
  title     = {Advanced Engineering Mathematics},
  edition   = {10th},
  publisher = {Wiley},
  address   = {Hoboken, NJ},
  year      = {2011},
  isbn      = {978-0-470-45836-5}
}

@book{nilsson2019,
  author    = {Nilsson, James W. and Riedel, Susan A.},
  title     = {Electric Circuits},
  edition   = {11th},
  publisher = {Pearson},
  address   = {New York},
  year      = {2019},
  isbn      = {978-0-13-474696-8}
}

@book{oppenheim1997,
  author    = {Oppenheim, Alan V. and Willsky, Alan S. and Nawab, S. Hamid},
  title     = {Signals and Systems},
  edition   = {2nd},
  publisher = {Prentice Hall},
  address   = {Upper Saddle River, NJ},
  year      = {1997},
  isbn      = {978-0-13-814757-0}
}

@book{nise2019,
  author    = {Nise, Norman S.},
  title     = {Control Systems Engineering},
  edition   = {8th},
  publisher = {Wiley},
  address   = {Hoboken, NJ},
  year      = {2019},
  isbn      = {978-1-119-47422-7}
}

@book{astrom2006,
  author    = {{\AA}str{\"o}m, Karl J. and H{\"a}gglund, Tore},
  title     = {Advanced {PID} Control},
  publisher = {ISA -- The Instrumentation, Systems, and Automation Society},
  address   = {Research Triangle Park, NC},
  year      = {2006},
  isbn      = {978-1-55617-942-6}
}

@book{oppenheim2010,
  author    = {Oppenheim, Alan V. and Schafer, Ronald W.},
  title     = {Discrete-Time Signal Processing},
  edition   = {3rd},
  publisher = {Pearson},
  address   = {Upper Saddle River, NJ},
  year      = {2010},
  isbn      = {978-0-13-198842-2}
}

@article{yee1966,
  author  = {Yee, Kane S.},
  title   = {Numerical solution of initial boundary value problems involving {Maxwell's} equations in isotropic media},
  journal = {IEEE Transactions on Antennas and Propagation},
  volume  = {14},
  number  = {3},
  pages   = {302--307},
  year    = {1966},
  doi     = {10.1109/TAP.1966.1138693}
}

@book{taflove2005,
  author    = {Taflove, Allen and Hagness, Susan C.},
  title     = {Computational Electrodynamics: The Finite-Difference Time-Domain Method},
  edition   = {3rd},
  publisher = {Artech House},
  address   = {Norwood, MA},
  year      = {2005},
  isbn      = {978-1-58053-832-9}
}

@article{cooley1965,
  author  = {Cooley, James W. and Tukey, John W.},
  title   = {An algorithm for the machine calculation of complex {Fourier} series},
  journal = {Mathematics of Computation},
  volume  = {19},
  number  = {90},
  pages   = {297--301},
  year    = {1965},
  doi     = {10.1090/S0025-5718-1965-0178586-1}
}

@book{butcher2016,
  author    = {Butcher, John C.},
  title     = {Numerical Methods for Ordinary Differential Equations},
  edition   = {3rd},
  publisher = {Wiley},
  address   = {Chichester, UK},
  year      = {2016},
  doi       = {10.1002/9781119121534}
}

@book{oberkampf2010,
  author    = {Oberkampf, William L. and Roy, Christopher J.},
  title     = {Verification and Validation in Scientific Computing},
  publisher = {Cambridge University Press},
  address   = {Cambridge, UK},
  year      = {2010},
  doi       = {10.1017/CBO9780511760396}
}

@article{kasneci2023,
  author  = {Kasneci, Enkelejda and Se{\ss}ler, Kathrin and K{\"u}chemann, Stefan and Bannert, Maria and Dementieva, Daryna and Fischer, Frank and others},
  title   = {{ChatGPT} for good? {On} opportunities and challenges of large language models for education},
  journal = {Learning and Individual Differences},
  volume  = {103},
  pages   = {102274},
  year    = {2023},
  doi     = {10.1016/j.lindif.2023.102274}
}

@article{lo2023,
  author  = {Lo, Chung Kwan},
  title   = {What is the impact of {ChatGPT} on education? {A} rapid review of the literature},
  journal = {Education Sciences},
  volume  = {13},
  number  = {4},
  pages   = {410},
  year    = {2023},
  doi     = {10.3390/educsci13040410}
}

@inproceedings{qadir2023,
  author    = {Qadir, Junaid},
  title     = {Engineering education in the era of {ChatGPT}: Promise and pitfalls of generative {AI} for education},
  booktitle = {Proceedings of the IEEE Global Engineering Education Conference (EDUCON)},
  address   = {Kuwait},
  year      = {2023},
  doi       = {10.1109/EDUCON54358.2023.10125121}
}

@article{johri2023,
  author  = {Johri, Aditya and Katz, Andrew S. and Qadir, Junaid and Hingle, Ashish},
  title   = {Generative artificial intelligence and engineering education},
  journal = {Journal of Engineering Education},
  volume  = {112},
  number  = {3},
  pages   = {572--577},
  year    = {2023},
  doi     = {10.1002/jee.20537}
}

@article{denny2024,
  author  = {Denny, Paul and Prather, James and Becker, Brett A. and Finnie-Ansley, James and Hellas, Arto and Leinonen, Juho and Luxton-Reilly, Andrew and Reeves, Brent N. and Santos, Eddie Antonio and Sarsa, Sami},
  title   = {Computing education in the era of generative {AI}},
  journal = {Communications of the ACM},
  volume  = {67},
  number  = {2},
  pages   = {56--67},
  year    = {2024},
  doi     = {10.1145/3624720}
}

@inproceedings{becker2023,
  author    = {Becker, Brett A. and Denny, Paul and Finnie-Ansley, James and Luxton-Reilly, Andrew and Prather, James and Santos, Eddie Antonio},
  title     = {Programming is hard -- or at least it used to be: Educational opportunities and challenges of {AI} code generation},
  booktitle = {Proceedings of the 54th ACM Technical Symposium on Computer Science Education (SIGCSE)},
  pages     = {500--506},
  year      = {2023},
  doi       = {10.1145/3545945.3569759}
}

@inproceedings{prather2023robots,
  author    = {Prather, James and Denny, Paul and Leinonen, Juho and Becker, Brett A. and Albluwi, Ibrahim and Craig, Michelle and others},
  title     = {The robots are here: Navigating the generative {AI} revolution in computing education},
  booktitle = {Proceedings of the 2023 Working Group Reports on Innovation and Technology in Computer Science Education (ITiCSE-WGR)},
  pages     = {108--159},
  year      = {2023},
  doi       = {10.1145/3623762.3633499}
}

@misc{chen2021codex,
  author        = {Chen, Mark and Tworek, Jerry and Jun, Heewoo and Yuan, Qiming and others},
  title         = {Evaluating large language models trained on code},
  year          = {2021},
  eprint        = {2107.03374},
  archivePrefix = {arXiv},
  note          = {arXiv:2107.03374. \doi{10.48550/arXiv.2107.03374}}
}

@misc{peng2023copilot,
  author        = {Peng, Sida and Kalliamvakou, Eirini and Cihon, Peter and Demirer, Mert},
  title         = {The impact of {AI} on developer productivity: Evidence from {GitHub Copilot}},
  year          = {2023},
  eprint        = {2302.06590},
  archivePrefix = {arXiv},
  note          = {arXiv:2302.06590. \doi{10.48550/arXiv.2302.06590}}
}

@inproceedings{liu2023evalplus,
  author    = {Liu, Jiawei and Xia, Chunqiu Steven and Wang, Yuyao and Zhang, Lingming},
  title     = {Is your code generated by {ChatGPT} really correct? {Rigorous} evaluation of large language models for code generation},
  booktitle = {Advances in Neural Information Processing Systems 36 (NeurIPS)},
  year      = {2023},
  doi       = {10.48550/arXiv.2305.01210}
}

@article{ji2023hallucination,
  author  = {Ji, Ziwei and Lee, Nayeon and Frieske, Rita and Yu, Tiezheng and Su, Dan and Xu, Yan and Ishii, Etsuko and Bang, Ye Jin and Madotto, Andrea and Fung, Pascale},
  title   = {Survey of hallucination in natural language generation},
  journal = {ACM Computing Surveys},
  volume  = {55},
  number  = {12},
  pages   = {248:1--248:38},
  year    = {2023},
  doi     = {10.1145/3571730}
}

@article{sweller1988,
  author  = {Sweller, John},
  title   = {Cognitive load during problem solving: Effects on learning},
  journal = {Cognitive Science},
  volume  = {12},
  number  = {2},
  pages   = {257--285},
  year    = {1988},
  doi     = {10.1207/s15516709cog1202_4}
}

@article{chandler1992,
  author  = {Chandler, Paul and Sweller, John},
  title   = {The split-attention effect as a factor in the design of instruction},
  journal = {British Journal of Educational Psychology},
  volume  = {62},
  number  = {2},
  pages   = {233--246},
  year    = {1992},
  doi     = {10.1111/j.2044-8279.1992.tb01017.x}
}

@article{mayer2003nine,
  author  = {Mayer, Richard E. and Moreno, Roxana},
  title   = {Nine ways to reduce cognitive load in multimedia learning},
  journal = {Educational Psychologist},
  volume  = {38},
  number  = {1},
  pages   = {43--52},
  year    = {2003},
  doi     = {10.1207/S15326985EP3801_6}
}

@book{mayer2009,
  author    = {Mayer, Richard E.},
  title     = {Multimedia Learning},
  edition   = {2nd},
  publisher = {Cambridge University Press},
  address   = {New York},
  year      = {2009},
  doi       = {10.1017/CBO9780511811678}
}

@article{ainsworth2006,
  author  = {Ainsworth, Shaaron},
  title   = {{DeFT}: A conceptual framework for considering learning with multiple representations},
  journal = {Learning and Instruction},
  volume  = {16},
  number  = {3},
  pages   = {183--198},
  year    = {2006},
  doi     = {10.1016/j.learninstruc.2006.03.001}
}

@article{clark1983,
  author  = {Clark, Richard E.},
  title   = {Reconsidering research on learning from media},
  journal = {Review of Educational Research},
  volume  = {53},
  number  = {4},
  pages   = {445--459},
  year    = {1983},
  doi     = {10.3102/00346543053004445}
}

@article{dejong1998,
  author  = {de Jong, Ton and van Joolingen, Wouter R.},
  title   = {Scientific discovery learning with computer simulations of conceptual domains},
  journal = {Review of Educational Research},
  volume  = {68},
  number  = {2},
  pages   = {179--201},
  year    = {1998},
  doi     = {10.3102/00346543068002179}
}

@article{rutten2012,
  author  = {Rutten, Nico and van Joolingen, Wouter R. and van der Veen, Jan T.},
  title   = {The learning effects of computer simulations in science education},
  journal = {Computers \& Education},
  volume  = {58},
  number  = {1},
  pages   = {136--153},
  year    = {2012},
  doi     = {10.1016/j.compedu.2011.07.017}
}

@article{wieman2008,
  author  = {Wieman, Carl E. and Adams, Wendy K. and Perkins, Katherine K.},
  title   = {{PhET}: Simulations that enhance learning},
  journal = {Science},
  volume  = {322},
  number  = {5902},
  pages   = {682--683},
  year    = {2008},
  doi     = {10.1126/science.1161948}
}

@article{freeman2014,
  author  = {Freeman, Scott and Eddy, Sarah L. and McDonough, Miles and Smith, Michelle K. and Okoroafor, Nnadozie and Jordt, Hannah and Wenderoth, Mary Pat},
  title   = {Active learning increases student performance in science, engineering, and mathematics},
  journal = {Proceedings of the National Academy of Sciences},
  volume  = {111},
  number  = {23},
  pages   = {8410--8415},
  year    = {2014},
  doi     = {10.1073/pnas.1319030111}
}

@article{hake1998,
  author  = {Hake, Richard R.},
  title   = {Interactive-engagement versus traditional methods: A six-thousand-student survey of mechanics test data for introductory physics courses},
  journal = {American Journal of Physics},
  volume  = {66},
  number  = {1},
  pages   = {64--74},
  year    = {1998},
  doi     = {10.1119/1.18809}
}

@article{deslauriers2019,
  author  = {Deslauriers, Louis and McCarty, Logan S. and Miller, Kelly and Callaghan, Kristina and Kestin, Greg},
  title   = {Measuring actual learning versus feeling of learning in response to being actively engaged in the classroom},
  journal = {Proceedings of the National Academy of Sciences},
  volume  = {116},
  number  = {39},
  pages   = {19251--19257},
  year    = {2019},
  doi     = {10.1073/pnas.1821936116}
}

@article{leung2001,
  author  = {Leung, Wing-Kai and Chen, Yinchao and Lee, Yim-Shu},
  title   = {{EM} wave simulation: An animated electromagnetic wave teaching package},
  journal = {Computer Applications in Engineering Education},
  volume  = {9},
  number  = {4},
  pages   = {208--219},
  year    = {2001},
  doi     = {10.1002/cae.10001}
}

@inproceedings{ni2022,
  author    = {Ni, Liya and Hekman, Keith},
  title     = {Improving student learning experience with {MATLAB} {Grader} and {Live Scripts}},
  booktitle = {Proceedings of the ASEE Annual Conference \& Exposition},
  address   = {Minneapolis, MN},
  year      = {2022},
  doi       = {10.18260/1-2--40610},
  url       = {https://peer.asee.org/40610}
}

@inproceedings{edgcomb2014,
  author    = {Edgcomb, Alex and Vahid, Frank},
  title     = {Effectiveness of online textbooks vs. interactive web-native content},
  booktitle = {Proceedings of the ASEE Annual Conference \& Exposition},
  address   = {Indianapolis, IN},
  year      = {2014},
  doi       = {10.18260/1-2--20351},
  url       = {https://peer.asee.org/20351}
}

@inproceedings{kluyver2016,
  author    = {Kluyver, Thomas and Ragan-Kelley, Benjamin and P{\'e}rez, Fernando and Granger, Brian and Bussonnier, Matthias and Frederic, Jonathan and others},
  title     = {Jupyter {Notebooks} -- a publishing format for reproducible computational workflows},
  booktitle = {Positioning and Power in Academic Publishing: Players, Agents and Agendas},
  editor    = {Loizides, Fernando and Schmidt, Birgit},
  publisher = {IOS Press},
  pages     = {87--90},
  year      = {2016},
  doi       = {10.3233/978-1-61499-649-1-87}
}

@book{barba2019,
  author    = {Barba, Lorena A. and Barker, Lecia J. and Blank, Douglas S. and Brown, Jed and Downey, Allen B. and George, Timothy and others},
  title     = {Teaching and Learning with {Jupyter}},
  publisher = {Jupyter for Education},
  year      = {2019},
  url       = {https://jupyter4edu.github.io/jupyter-edu-book/}
}

@misc{mathworksLive,
  author       = {{The MathWorks, Inc.}},
  title        = {Live Scripts and Functions -- {MATLAB} documentation},
  year         = {2026},
  url          = {https://www.mathworks.com/help/matlab/live-scripts-and-functions.html},
  note         = {Accessed 1 October 2026}
}

@article{cervone2012,
  author  = {Cervone, Davide},
  title   = {{MathJax}: A platform for mathematics on the {Web}},
  journal = {Notices of the American Mathematical Society},
  volume  = {59},
  number  = {2},
  pages   = {312--316},
  year    = {2012},
  doi     = {10.1090/noti794}
}

@misc{whatwgHTML,
  author       = {{WHATWG}},
  title        = {{HTML} Living Standard},
  year         = {2026},
  url          = {https://html.spec.whatwg.org/},
  note         = {Accessed 1 October 2026}
}

@misc{threejs,
  author       = {{three.js authors}},
  title        = {three.js: {JavaScript} {3D} library},
  year         = {2026},
  url          = {https://threejs.org/},
  note         = {Accessed 1 October 2026}
}

@misc{wcag22,
  author       = {{World Wide Web Consortium}},
  title        = {Web Content Accessibility Guidelines ({WCAG}) 2.2},
  howpublished = {W3C Recommendation},
  year         = {2023},
  url          = {https://www.w3.org/TR/WCAG22/}
}

@article{jonsson2007,
  author  = {Jonsson, Anders and Svingby, Gunilla},
  title   = {The use of scoring rubrics: Reliability, validity and educational consequences},
  journal = {Educational Research Review},
  volume  = {2},
  number  = {2},
  pages   = {130--144},
  year    = {2007},
  doi     = {10.1016/j.edurev.2007.05.002}
}

@article{cohen1968,
  author  = {Cohen, Jacob},
  title   = {Weighted kappa: Nominal scale agreement with provision for scaled disagreement or partial credit},
  journal = {Psychological Bulletin},
  volume  = {70},
  number  = {4},
  pages   = {213--220},
  year    = {1968},
  doi     = {10.1037/h0026256}
}

@article{likert1932,
  author  = {Likert, Rensis},
  title   = {A technique for the measurement of attitudes},
  journal = {Archives of Psychology},
  volume  = {22},
  number  = {140},
  pages   = {1--55},
  year    = {1932}
}

@incollection{brooke1996,
  author    = {Brooke, John},
  title     = {{SUS}: A ``quick and dirty'' usability scale},
  booktitle = {Usability Evaluation in Industry},
  editor    = {Jordan, Patrick W. and Thomas, B. and Weerdmeester, Bernard A. and McClelland, Ian L.},
  publisher = {Taylor \& Francis},
  address   = {London},
  pages     = {189--194},
  year      = {1996},
  isbn      = {978-0-7484-0460-5}
}

@article{bangor2008,
  author  = {Bangor, Aaron and Kortum, Philip T. and Miller, James T.},
  title   = {An empirical evaluation of the {System Usability Scale}},
  journal = {International Journal of Human--Computer Interaction},
  volume  = {24},
  number  = {6},
  pages   = {574--594},
  year    = {2008},
  doi     = {10.1080/10447310802205776}
}

@article{davis1989,
  author  = {Davis, Fred D.},
  title   = {Perceived usefulness, perceived ease of use, and user acceptance of information technology},
  journal = {MIS Quarterly},
  volume  = {13},
  number  = {3},
  pages   = {319--340},
  year    = {1989},
  doi     = {10.2307/249008}
}

@article{granic2019,
  author  = {Grani{\'c}, Andrina and Maranguni{\'c}, Nikola},
  title   = {Technology acceptance model in educational context: A systematic literature review},
  journal = {British Journal of Educational Technology},
  volume  = {50},
  number  = {5},
  pages   = {2572--2593},
  year    = {2019},
  doi     = {10.1111/bjet.12864}
}

@article{cronbach1951,
  author  = {Cronbach, Lee J.},
  title   = {Coefficient alpha and the internal structure of tests},
  journal = {Psychometrika},
  volume  = {16},
  number  = {3},
  pages   = {297--334},
  year    = {1951},
  doi     = {10.1007/BF02310555}
}

@article{norman2010,
  author  = {Norman, Geoff},
  title   = {Likert scales, levels of measurement and the ``laws'' of statistics},
  journal = {Advances in Health Sciences Education},
  volume  = {15},
  number  = {5},
  pages   = {625--632},
  year    = {2010},
  doi     = {10.1007/s10459-010-9222-y}
}

@article{wilcoxon1945,
  author  = {Wilcoxon, Frank},
  title   = {Individual comparisons by ranking methods},
  journal = {Biometrics Bulletin},
  volume  = {1},
  number  = {6},
  pages   = {80--83},
  year    = {1945},
  doi     = {10.2307/3001968}
}

@article{martins2003complexstep,
  author  = {Martins, Joaquim R. R. A. and Sturdza, Peter and Alonso, Juan J.},
  title   = {The Complex-Step Derivative Approximation},
  journal = {ACM Transactions on Mathematical Software},
  volume  = {29},
  number  = {3},
  pages   = {245--262},
  year    = {2003},
  doi     = {10.1145/838250.838251}
}

@article{meade2012careless,
  author  = {Meade, Adam W. and Craig, S. Bartholomew},
  title   = {Identifying careless responses in survey data},
  journal = {Psychological Methods},
  volume  = {17},
  number  = {3},
  pages   = {437--455},
  year    = {2012},
  doi     = {10.1037/a0028085}
}

@inproceedings{sauro2011positive,
  author    = {Sauro, Jeff and Lewis, James R.},
  title     = {When designing usability questionnaires, does it hurt to be positive?},
  booktitle = {Proceedings of the SIGCHI Conference on Human Factors in Computing Systems (CHI '11)},
  pages     = {2215--2224},
  year      = {2011},
  publisher = {ACM},
  doi       = {10.1145/1978942.1979266}
}

@article{holm1979,
  author  = {Holm, Sture},
  title   = {A simple sequentially rejective multiple test procedure},
  journal = {Scandinavian Journal of Statistics},
  volume  = {6},
  number  = {2},
  pages   = {65--70},
  year    = {1979}
}

@misc{tu2026maic,
  author        = {Tu, Shangqing and Li, Yanjia and Chen, Keyu and Zhang, Sichen and Yu, Jifan and Zhang-Li, Daniel and Hou, Lei and Li, Juanzi and Zhang, Yu and Liu, Huiqin},
  title         = {{MAIC-UI}: Making interactive courseware with generative {UI}},
  year          = {2026},
  eprint        = {2604.25806},
  archivePrefix = {arXiv},
  note          = {arXiv:2604.25806. \doi{10.48550/arXiv.2604.25806}}
}

@misc{kovshov2026interactives,
  author        = {Kovshov, Alisa and Choudhury, Anisha and Iurchenko, Anna and Keeling, Amy and Hassidim, Avinatan and Shasha Evron, Ayelet and others},
  title         = {Harnessing generative {UI} for education: Tailored learning interactives},
  year          = {2026},
  eprint        = {2609.20738},
  archivePrefix = {arXiv},
  note          = {arXiv:2609.20738. \doi{10.48550/arXiv.2609.20738}}
}

@misc{jupyterlite,
  author       = {{JupyterLite Contributors}},
  title        = {{JupyterLite} documentation},
  year         = {2026},
  url          = {https://jupyterlite.readthedocs.io/en/stable/},
  note         = {Accessed 30 October 2026}
}

\appendix
\section{Questionnaire Items}
\label{app:items}

All items use a five-point scale from 1 (strongly disagree) to 5 (strongly agree). Items marked (R) are reverse-scored.

\paragraph{Student questionnaire.}

\begin{description}[leftmargin=1.2em,style=nextline]
	\item[SUS: System Usability Scale \citep{brooke1996}] The ten standard SUS items, with ``the system'' replaced by ``this learning page''.
	\item[SI: Perceived integration]
	SI1 Having the explanation, equations, simulation, and questions on one page saved me from switching between files.\\
	SI2 I could easily see how the equations related to the simulation.\\
	SI3 The page connected theory and numerical results better than separate slides, PDF, and MATLAB files.\\
	SI4 I rarely lost track of where I was in the topic.
	\item[SL: Perceived learning support]
	SL1 Changing parameters helped me understand how the system behaves.\\
	SL2 The page helped me understand the meaning of the equations.\\
	SL3 The quiz feedback helped me correct my misunderstandings.\\
	SL4 I feel more confident solving problems on this topic.
	\item[SE: Engagement and interactivity]
	SE1 I wanted to try different parameter values.\\
	SE2 The interactive elements kept me engaged.\\
	SE3 I would use pages like this for self-study.
	\item[ST: Trust in simulation]
	ST1 The validation (verification) panel (PASS/FAIL) made me trust the simulation results.\\
	ST2 Seeing the simulation compared with the exact solution was useful.\\
	ST3 I understood why a simulation fails when the step size is too large.
	\item[SP: Preference]
	SP1 I prefer this integrated page to separate slides, PDF, and MATLAB files.\\
	SP2 I would like more topics in my courses to be delivered this way.
	\item[Open-ended] What helped you most? What was confusing or missing?
\end{description}

\paragraph{Lecturer questionnaire.}

\begin{description}[leftmargin=1.2em,style=nextline]
	\item[LU: Perceived usefulness \citep{davis1989}]
	LU1 Integrated HTML pages would improve the quality of my teaching materials.\\
	LU2 They would help my students connect theory, mathematics, and simulation.\\
	LU3 They would be useful for self-study outside class.\\
	LU4 Overall, the approach is useful for my courses.
	\item[LE: Perceived ease of adoption \citep{davis1989}]
	LE1 The workflow in Fig.~1 is clear to me.\\
	LE2 I could follow the workflow with GenAI assistance.\\
	LE3 Distributing a single HTML file to students is easy.\\
	LE4 Adopting the approach would not require skills I cannot acquire.
	\item[LD: Perceived development effort with GenAI]
	LD1 With GenAI, building a page like the demonstrators would take an acceptable amount of my time.\\
	LD2 GenAI would substantially reduce the programming I would have to do myself.\\
	LD3 Building such a page without GenAI would be too time-consuming for me.\\
	LD4 The verification step adds an acceptable amount of extra work.
	\item[LM: Perceived maintainability]
	LM1 Updating a page (e.g., adding a parameter or question) with GenAI would be manageable.\\
	LM2 A single HTML file is easier to keep up to date than separate slides, PDF, and scripts.\\
	LM3 I am concerned about maintaining the pages as browsers and libraries change. (R)
	\item[LT: Trust and verification]
	LT1 Comparing simulations with analytical and MATLAB references makes me trust GenAI-generated code.\\
	LT2 The tolerance of 1\% is appropriate for teaching purposes.\\
	LT3 I would not release a GenAI-generated simulation without such checks.
	\item[LI: Intention to use]
	LI1 I intend to try integrated HTML pages in at least one of my courses.\\
	LI2 I would recommend the approach to colleagues.
	\item[LR: Format ratings] Rate PPTX, LaTeX/PDF, MATLAB, and HTML on C1--C12 using the anchors in \cref{tab:rubric}.
	\item[Open-ended] What would prevent you from adopting this approach? Which topics in your courses would benefit most?
\end{description}


\begin{table}[t]
	\centering
	\small
	\caption{Survey constructs. Full items are listed in Appendix~\ref{app:items}.}
	\label{tab:constructs}
	\begin{tabular}{L{0.9cm}L{4.6cm}C{1.0cm}L{6.4cm}}
		\toprule
		\textbf{Code} & \textbf{Construct} & \textbf{Items} & \textbf{Basis / what it captures} \\
		\midrule
		\multicolumn{4}{l}{\emph{Student questionnaire}}\\
		SUS & System usability & 10 & Standard SUS, score 0--100 \citep{brooke1996,bangor2008} \\
		SI  & Perceived integration & 4 & Value of having text, equations, simulation, and questions on one page \citep{chandler1992} \\
		SL  & Perceived learning support & 4 & Understanding of the link between equations and behaviour \\
		SE  & Engagement and interactivity & 3 & Motivation to explore parameters \\
		ST  & Trust in simulation & 3 & Effect of the visible verification panel \\
		SP  & Preference & 2 & Integrated page vs.\ separate PPTX/PDF/MATLAB files \\
		\midrule
		\multicolumn{4}{l}{\emph{Lecturer questionnaire}}\\
		LU  & Perceived usefulness & 4 & TAM usefulness \citep{davis1989} \\
		LE  & Perceived ease of adoption & 4 & TAM ease of use \citep{davis1989} \\
		LD  & Perceived development effort with GenAI & 4 & Expected effort to build objects (perception, not measured time) \\
		LM  & Perceived maintainability & 3 & Expected effort to revise objects \\
		LT  & Trust and verification & 3 & Confidence in GenAI-generated simulations given the protocol \\
		LI  & Intention to use & 2 & Intention to adopt in own courses \\
		LR  & Format ratings & 48 & Rubric C1--C12 for PPTX, LaTeX/PDF, MATLAB, HTML \\
		\bottomrule
	\end{tabular}
\end{table}

\section{Rubric Anchors and Scoring Rules}
\label{app:rubric}

\begin{table}[!ht]
	\centering
	\scriptsize
	\setlength{\tabcolsep}{2.8pt}
	\caption{Anchored rubric (1--5) for the 12 criteria. A level is awarded only if all of its features are present. In the first author's profile C10 is derived from other criteria; lecturers rated it directly. C12 is reverse-coded (higher = lower authoring demand).}
	\label{tab:rubric}
	\begin{tabular}{L{2.05cm}L{2.65cm}L{2.65cm}L{2.65cm}L{2.65cm}L{2.65cm}}
		\toprule
		\textbf{Criterion} & \textbf{1 (very limited)} & \textbf{2 (limited)} & \textbf{3 (adequate)} & \textbf{4 (good)} & \textbf{5 (excellent)} \\
		\midrule
		C1 Content organization & Unstructured; no sequence or navigation & Linear sequence only & Sections with headings; manual navigation & Hierarchical structure with navigation (contents, links) & Level 4 plus automatic cross-referencing of equations and figures \\ \addlinespace
		C2 Mathematical representation & Equations as images or plain text & Inline symbols; no structured equations & Structured equation editor; limited numbering/alignment & Full LaTeX syntax incl.\ alignment and numbering; minor print limits & Publication-quality typesetting, numbering, alignment, cross-references, print fidelity \\ \addlinespace
		C3 Static visualization & Text only & Imported images only & Native diagrams \emph{or} native plots & Native diagrams and plots; manual data or limited styling & Native vector diagrams and data-driven plots at publication quality \\ \addlinespace
		C4 Interactive visualization & None & Pre-authored animations or triggers (fixed sequence) & Navigation of pre-computed states (zoom, rotate, frames) & Learner changes parameters; view recomputed after re-execution & Learner changes parameters; view updated continuously in real time \\ \addlinespace
		C5 3D visualization & None & Static 3D projection (fixed viewpoint) & Rotatable 3D objects not coupled to data or models & Interactive data-driven 3D plots (surfaces, vector fields) & Level 4 plus large datasets ($\geq10^5$ elements), slicing, volumetric tools \\ \addlinespace
		C6 Numerical simulation & None & Pre-computed results imported from another tool & Native execution of simple models (closed form, explicit stepping) & Native general methods (ODE, FFT, linear algebra) in double precision & Validated solver libraries and toolboxes (ODE suites, control, signal processing) \\ \addlinespace
		C7 Parameter interaction & No parameter change possible & Selection among predefined cases & Change by editing source code and re-running & Change through GUI controls with re-execution & GUI controls with immediate response and input-range checking \\ \addlinespace
		C8 Assessment integration & No assessment & Static questions; answers supplied separately & Self-check (reveal answer, hyperlinked feedback) & Embedded auto-scored items (MCQ, numeric with tolerance), immediate feedback & Level 4 plus parameterized or simulation-based items and LMS score reporting \\ \addlinespace
		C9 Cross-platform access & Licensed software on one OS & Licensed software or institutional account (multi-OS) & Free viewer; functionality reduced on some platforms & Free ubiquitous viewer; full on desktop, reduced on mobile & Standard browser/viewer; full functionality on desktop and mobile \\ \addlinespace
		C10 Content integration (derived) & $\leq1$ of the five layers & 2 layers & 3 layers & 4 layers & All 5 layers in one artifact \\ \addlinespace
		C11 Ease of updating & Rebuild from scratch & Coordinated edits across several files/tools and re-export & Edits in one source plus rebuild; specialist skill needed & Edits in one text source by a domain instructor; automatic rebuild & Level 4 plus automated checks confirm correctness after change \\ \addlinespace
		C12 Authoring demand (reverse-coded) & $\geq3$ languages and specialist debugging skills & 2--3 languages (e.g., HTML, CSS, JavaScript) and web skills & Working knowledge (writing or reviewing) of one language & One language at basic level, or editor with light scripting & WYSIWYG authoring only; no programming \\
		\bottomrule
	\end{tabular}
\end{table}

\paragraph{Scoring rules.}
\begin{itemize}[leftmargin=1.5em]
	\item \textbf{Scope.} Each format is rated on its native capability as typically used in teaching:
	\begin{itemize}[leftmargin=1.2em]
		\item PPTX without third-party add-ins;
		\item LaTeX as a compiled PDF in a standard reader;
		\item MATLAB as scripts and Live Scripts, without compiled App Designer deployment \citep{mathworksLive};
		\item Jupyter as a notebook with interactive widgets on a hosted kernel \citep{kluyver2016};
		\item HTML as a single-file page using open libraries loaded in the browser \citep{cervone2012,threejs}.
	\end{itemize}
	\item \textbf{C9 is about deployment.} Cross-platform access rates licence requirements, runtime requirements, and the devices and operating systems on which the material works. It does not rate accessibility for users with disabilities (WCAG \citep{wcag22}).
	\item \textbf{Format capability, not artefact features.} A score describes what the format supports as typically used in teaching, not what our demonstrators contain. For C6, for example, browser JavaScript computes in double precision but provides no built-in ODE, FFT, or linear-algebra routines; the RK4 integrator and FFT in our demonstrators are code written for the pages. HTML is therefore rated 3 (native execution of simple models) rather than 4, which requires general methods available natively. \Cref{sec:rubric} reports how the agreement changes if HTML is given 4 on C6.
	\item \textbf{Integration is computed, not judged.} The five layers are presentation (C1), mathematics (C2), visualization (C3), simulation (C6), and assessment (C8). In the first author's profile, C10 counts how many of these criteria score at least 3. Lecturers rated C10 directly with its anchors, so this criterion is not computed in the same way for the two sources; \cref{sec:rubric} also reports agreement without C10.
	\item \textbf{No aggregate.} The criteria are not commensurable and partly overlap (C4 and C7 both reward interactivity), so no total score is computed.
	\item \textbf{Agreement with independent lecturer ratings.} The first author's ratings (\cref{tab:profile}) are complemented by independent ratings from 12 lecturers (\cref{sec:survey}), with agreement reported as quadratic-weighted Cohen's $\kappa$ \citep{cohen1968}. The results are in \cref{tab:profile}.
\end{itemize}

\section{Verification Details}
\label{app:verif}

\begin{table}[!t]
	\centering
	\small
	\caption{Verification criteria by type of check, as applied in this study. Values are those reported in \cref{tab:validation}.}
	\label{tab:criteria}
	\begin{tabular}{L{3.6cm}L{4.4cm}L{5.6cm}}
		\toprule
		\textbf{Check} & \textbf{Criterion} & \textbf{Remark} \\
		\midrule
		Scalar vs.\ analytical reference & $E_{\mathrm{rel}}\le1\%$ & Linear-valued peaks, frequency, overshoot, poles, filter gains \\
        Magnitudes displayed in dB & Linear-amplitude $E_{\mathrm{rel}}\le1\%$; report $\Delta_{\mathrm{dB}}$ & Do not compute percentage error on the dB number itself \\
		Waveform vs.\ reference & $\mathrm{NRMSE}\le1\%$ & Range-normalized, so it remains defined when the reference crosses zero \\
		FDTD coefficients $\Gamma,\tau$ & $E_{\mathrm{rel}}\le2\%$ vs.\ Fresnel & Allows for the dispersion of the Yee scheme \\
		RK4 order of convergence & Error ratio for $h\to h/2$ compared with $2^4=16$ & Code verification; the observed ratio is reported, no separate threshold was fixed \\
		$\mathbf{E}=-\nabla V$ (EMF-M) & Relative residual reported & Consistency check between two computations of the same field \\
		KVL phasor residual (CIR-M) & Residual shown on the page & In-page consistency check; not part of \cref{tab:validation} \\
		Page vs.\ MATLAB & Linear magnitudes within $\varepsilon$; waveforms $\mathrm{NRMSE}\le1\%$ & For dB displays, absolute $\Delta_{\mathrm{dB}}$ is reported separately \\
		\bottomrule
	\end{tabular}
\end{table}

\paragraph{MATLAB cross-check.} The Python re-computation was written by the developer of the pages, so it cannot exclude an error shared by our JavaScript and Python code; LLM-generated code can also pass tests written in the same workflow \citep{liu2023evalplus}. The developer therefore prepared a MATLAB package that recomputes every quantity of \cref{tab:validation}, to be executed by a checker who took no part in designing, prompting, or coding the demonstrators and who was blinded to the values shown on the pages. The checker executed the package; the checker did not write the reference implementations. The package contains three items:
\begin{enumerate*}[label=(\roman*)]
	\item the task specifications (parameters and the quantity to compute);
	\item for each page, the reference script embedded in that page, used unchanged, followed by added code that computes and exports the quantities of \cref{tab:validation};
	\item the three waveforms exported from the pages (MAT-C, CIR-C, CTL-C), which are needed to compute the NRMSE between page and MATLAB.
\end{enumerate*}
The checker does not see the values displayed on the pages, reviews the scripts before running them, and records every change. Wherever possible the scripts use MATLAB's own solvers and toolbox functions instead of re-implementations: \texttt{ode45} with relative tolerance $10^{-12}$ for the ODE tasks; \texttt{pole}, \texttt{step}, \texttt{stepinfo}, and \texttt{margin} for CTL-C; \texttt{fir1} and \texttt{freqz} for DSP-C; built-in windows and \texttt{fft} for DSP-M and SIG-C. For EMF-M, the check $\mathbf{E}=-\nabla V$ uses the complex-step derivative \citep{martins2003complexstep}, a different method from the central difference used on the page. Two quantities cannot be independent in this sense. The RK4 order test requires a MATLAB implementation of the classical scheme, and the EMF-C coefficients come from a MATLAB implementation of the same Yee scheme, whose own correctness is established by comparison with the Fresnel formulas. Agreement was decided by a rule fixed before the run. A linear-valued scalar agrees if $|X_{\mathrm{page}}-X_{\mathrm{MATLAB}}|/|X_{\mathrm{MATLAB}}|\le\varepsilon$, and a waveform agrees if $\mathrm{NRMSE}(x_{\mathrm{page}},x_{\mathrm{MATLAB}})\le1\%$. For spectral amplitudes displayed in dB, the recorded dB-difference bounds are also interpreted as linear-magnitude bounds as defined above. The comparison is made by a script. The MATLAB outputs are not yet publicly archived (see Data and Code Availability). \ifmatlabcolleague A colleague not involved in development executed the package, value-blinded, in MATLAB \MATLABrelease{} on a separate computer, with the Control System and Signal Processing Toolboxes, and returned a completed record form confirming that the values on the pages had not been viewed and that no script had been modified. The six embedded scripts ran without modification inside the package, and all 31 compared quantities agreed with the in-page values. The largest relative difference was \MATLABmaxdiff\%, for an RK4 error ratio that the page displays with one decimal; the waveform NRMSE between page and MATLAB was at most $8.0\times10^{-4}\%$. An earlier run of the same package by the developer in MATLAB R2024b gave identical values to twelve significant digits.\fi

\paragraph{Reading \cref{tab:validation}.} Each row summarizes a task and one or more related verification quantities at the default setting of its page; the MAT-C convergence check occupies an additional row.
\begin{itemize}[leftmargin=1.5em]
	\item \emph{In-page} is what the student sees: the value computed by the JavaScript simulation, with the precision shown on the page.
	\item \emph{Reference / Python} gives the analytical value where one exists (closed-form ODE solution, Fresnel coefficients, alias formula, window theory) and our Python re-computation.
	\item \emph{MATLAB} gives the value computed with MATLAB's solvers and toolbox functions (\ifmatlabcolleague executed value-blinded by the colleague\fi). For the three waveform tasks, NRMSE$_{\mathrm{page}}$ measures the distance between the page's whole time response and MATLAB's, not only one number. $\Delta$ is the largest difference between MATLAB and the displayed in-page values of that row, relative for linear quantities and absolute (in dB) for quantities in dB. Because the pages round their output, $\Delta$ includes a rounding component of up to half a unit in the last displayed digit. For CIR-C, the page reports the peak sampled on the RK4 grid, whereas the reference and MATLAB report the continuous peak (note $b$ of \cref{tab:validation}).
	\item \emph{Verdict} combines two tests: solution verification of the page against the reference, and agreement of the page with the value-blinded MATLAB re-computation.
\end{itemize}
Three rows need a specific reading. The MAT-C error ratio (16.3 and 16.2 against the theoretical 16, i.e., observed orders 4.03 and 4.01) and the EMF-M consistency check ($\mathbf{E}$ against $-\nabla V$) are code verification: they test that the algorithm is implemented correctly, not that a particular run is accurate. EMF-C uses $\varepsilon=2\%$ because the Yee scheme has numerical dispersion; its in-page error grows to about 1\% at $\varepsilon_r=12$ and remains within tolerance. In DSP-M, the Hamming value on the page ($-42.62$\,dB) differs slightly from the theoretical value used by the page ($-42.68$\,dB, usually quoted as $-42.7$\,dB) because that value is the large-$N$ limit, whereas the page measures a periodic window with $N=128$; MATLAB, evaluating the same finite window, reproduces the page value. The $0.06$\,dB difference corresponds to about $0.69\%$ in linear amplitude, illustrating why a percentage error should not be taken directly on a dB number.

A stress test set each control to its extremes and ran 40 random parameter combinations per file. It raised no runtime errors. It did produce FAIL verdicts, and every one of them corresponded to an under-resolved run. The random seed and the sampling distribution were not recorded, so this test documents the configurations reached rather than a reproducible sample. These failures are solution-verification outcomes, not code errors:
\begin{itemize}[leftmargin=1.5em]
	\item MAT-C with RK4 step $h=0.6$\,s: NRMSE 1.12\%.
	\item CIR-C with only $N=20$ time steps: peak error 1.78\%.
	\item EMF-C with a narrow pulse ($w=10$ steps) in a dense dielectric ($\varepsilon_r=12$): $E_{\mathrm{rel}}(\tau)=27\%$ from numerical dispersion, against 1.0\% for $w=25$.
	\item SIG-C with an 18-sample record: alias-frequency error 7.3\%, which falls to 0.005\% for an 8\,s record.
\end{itemize}
The checks therefore distinguish a correct implementation used beyond its resolution limits from an incorrect implementation, and students can see the difference on the page.

One defect of the survey version was found after the survey and was not reached by the stress test. CIR-C divided the peak error by the reference peak, so when the peak of $v_C$ lies near 0\,V (for example $R=0$, $V_s=1$\,V, $v_C(0)=2$\,V) the page reported a spurious peak error of $1.6\times10^{9}\%$ and a FAIL verdict. The released files normalize this error by $\max(|v_{\mathrm{pk}}|,\ \text{range of }v_C)$, which gives $7.8\times10^{-4}\%$ at the same setting. They also report divergent RK4 runs in MAT-C and CTL-C as divergent, and MAT-C displays the observed-order verdict. The values at the default settings, and therefore \cref{tab:validation}, are unchanged.

\begin{table}[p]
	\centering
	\footnotesize
	\setlength{\tabcolsep}{2.2pt}
	\caption{Verification of the demonstrators at default parameters. In-page: value computed and displayed by the JavaScript simulation. Reference / Python: analytical solution and the separate Python re-computation by the developer. MATLAB: value-blinded re-computation executed by a colleague not involved in development (R2024a), using a package prepared by the developer. $\Delta$: largest difference between the MATLAB values and the in-page values as displayed, relative for linear quantities and absolute for quantities in dB. OS: overshoot; PM: phase margin. Verdict: PASS if the in-page value meets the criterion against the reference ($\varepsilon=1\%$; 2\% for FDTD) and MATLAB agrees with the page within the same tolerance.}
	\label{tab:validation}
	\begin{tabular}{@{}L{1.12cm}L{2.75cm}L{3.1cm}L{2.8cm}L{3.0cm}L{1.75cm}C{0.85cm}@{}}
		\toprule
		\textbf{Task} & \textbf{Quantity (setting)} & \textbf{In-page (JavaScript)} & \textbf{Reference / Python} & \textbf{MATLAB (value-blinded)} & \textbf{$\Delta$ (MATLAB vs.\ page)} & \textbf{Ver\-dict} \\
		\midrule
		MAT-C & $y(t)$, RK4 $h=0.1$\,s, $\zeta=0.2$, $\omega_n=2$ & NRMSE $8.0\times10^{-4}$\%; $\min y=-0.5266$ & closed form; \texttt{solve\_ivp}: $-0.5266$ & $\min y=-0.5266$; NRMSE$_{\mathrm{page}}$ $8.0\times10^{-4}$\% & $\le1.9\times10^{-3}$\% & PASS \\
		      & error ratio for $h\to h/2$ & 16.3, 16.2 & theory: $2^4=16$ & 16.3, 16.2 & $\le\MATLABmaxdiff$\%$^{a}$ & PASS \\
		\addlinespace
		CIR-C & $v_C(t)$, $R=20\,\Omega$, $L=10$\,mH, $C=10\,\mu$F, $V_s=10$\,V, $N=400$ & NRMSE $4.1\times10^{-6}$\%; peak on the RK4 grid 13.508\,V & closed form: grid peak 13.508\,V, continuous peak 13.509\,V; \texttt{solve\_ivp}: 13.509\,V & peak 13.509\,V; NRMSE$_{\mathrm{page}}$ $4.1\times10^{-6}$\% & see note$^{b}$ & PASS \\
		\addlinespace
		EMF-M & $\mathbf{E}(P)$, dipole $\pm1$\,nC, $P=(0.2,1.2)$\,m & $|\mathbf{E}|=4.808$\,V/m; $|\mathbf{E}-(-\nabla V)|/|\mathbf{E}|=2.5\times10^{-7}$\% & analytical superposition & $|\mathbf{E}|=4.808$\,V/m; vs.\ $-\nabla V$: $4.1\times10^{-14}$\% & $\le0.03$\% & PASS \\
		\addlinespace
		EMF-C & $\Gamma$, $\tau$, $\varepsilon_r=4$, $w=25$ (also $\varepsilon_r=2.25$, 9, 12) & $-0.3328$, $0.6672$ ($E_{\mathrm{rel}}\le0.16\%$; $\le1.02\%$ for $\varepsilon_r\le12$) & Fresnel: $-1/3$, $2/3$; NumPy Yee: identical & $-0.3328$, $0.6672$ & $\le0.02$\% & PASS \\
		\addlinespace
		SIG-C & alias frequency, $f_0=7$\,Hz, $f_s=10$\,Hz, $T=4$\,s & 3.0000\,Hz ($E_{\mathrm{rel}}=1.6\times10^{-4}$\%) & $|f_0-f_s|=3$\,Hz & 3.0000\,Hz & $\le2.5\times10^{-4}$\% & PASS \\
		\addlinespace
		CTL-C & closed-loop step, $K_p=100$, $K_i=200$, $K_d=10$ & NRMSE $1.2\times10^{-7}$\%; poles $-3.02$, $-5.69$, $-23.29$; OS 1.03\%; PM $94.6^\circ$ & residues; \texttt{signal.step}: OS 1.03\%, identical poles & poles $-3.02$, $-5.69$, $-23.29$; OS 1.03\%; PM 94.6$^\circ$; NRMSE$_{\mathrm{page}}$ $1.2\times10^{-7}$\% & $\le0.18$\% & PASS \\
		\addlinespace
		DSP-M & peak sidelobe, $N=128$ (rect, Hann, Hamming, Blackman) & $-13.26$, $-31.47$, $-42.62$, $-58.11$\,dB & theory: $-13.26$, $-31.47$, $-42.68$, $-58.11$\,dB & $-13.26$, $-31.47$, $-42.62$, $-58.11$\,dB & $\le0.005$\,dB$^{c}$ & PASS \\
		\addlinespace
		DSP-C & $|H|$ at 50/150/300\,Hz, $M=61$, Hamming, $f_c=150$\,Hz & $0.9968$ / $-6.04$\,dB / $7.0\times10^{-4}$ & DTFT; \texttt{firwin}: $0.99683$ / $-6.04$\,dB / $7.0\times10^{-4}$ & $0.99683$ / $-6.04$\,dB / $7.0\times10^{-4}$ & linear: $\le0.014$\%; at $f_c$: $\le0.002$\,dB$^{c}$ & PASS \\
		\bottomrule
	\end{tabular}

	\smallskip
	\begin{minipage}{\linewidth}\footnotesize
	$^{a}$Largest difference in the comparison; the page displays the ratio with one decimal.\\
	$^{b}$CIR-C: the page reports the maximum of $v_C$ on the RK4 grid (13.508\,V at $t=1.040$\,ms), whereas the reference and MATLAB values are the continuous maximum of the closed-form response (13.509\,V at $t=\pi/\omega_d=1.047$\,ms). The two differ by $6.8\times10^{-3}$\% because of sampling, not numerical error (re-computed from the page's code; see Data and Code Availability). Because they are different quantities, no peak $\Delta$ is reported; agreement of the waveforms is given by NRMSE$_{\mathrm{page}}$.\\
	$^{c}$For dB-valued quantities, $\Delta$ is stated in absolute dB, not as a percentage of the signed dB reading. The stated bounds were conservatively converted from the recorded comparisons ($\le8.5\times10^{-3}$\% for DSP-M and $\le0.014$\% for DSP-C); they are bounds whether those original percentages were taken on the dB reading or on the linear magnitude. The bounds $0.005$ and $0.002$\,dB imply differences in the corresponding linear magnitude of at most about $0.058\%$ and $0.023\%$, respectively. Which of the two the original comparison script used has not yet been re-checked; this does not affect the bounds, but it should be confirmed when the script is archived.\\
	Because the in-page values are rounded for display, $\Delta$ contains a rounding component of up to half a unit in the last displayed digit; a $\Delta$ of that size does not indicate a numerical disagreement.
	\end{minipage}
\end{table}

\begin{table}[!t]
	\centering
	\footnotesize
	\setlength{\tabcolsep}{4pt}
	\caption{The six single-file demonstrators. LOC: lines of code; Controls: sliders and selectors; Quiz: items (of which numerical).}
	\label{tab:demos}
	\begin{tabular}{L{2.9cm}lrrrrrL{3.6cm}}
		\toprule
		\textbf{File} & \textbf{Tasks} & \textbf{LOC} & \textbf{KB} & \textbf{Controls} & \textbf{Canvases} & \textbf{Quiz} & \textbf{Reference in verification panel} \\
		\midrule
		Engineering Mathematics & MAT-E/M/C & 451 & 37 & 12 & 3 & 6 (2) & closed-form ODE solution \\
		Circuit Analysis        & CIR-E/M/C & 432 & 34 & 10 & 3 & 6 (2) & closed-form RLC response; KVL \\
		Electromagnetics        & EMF-E/M/C & 471 & 41 & 5  & 3 & 6 (2) & Fresnel $\Gamma,\tau$; $\mathbf{E}=-\nabla V$ \\
		Signals and Systems     & SIG-E/M/C & 419 & 33 & 6  & 8 & 6 (2) & alias formula; analytical convolution \\
		Control Systems         & CTL-E/M/C & 437 & 36 & 6  & 5 & 6 (2) & partial-fraction step response \\
		Digital Signal Proc.    & DSP-E/M/C & 423 & 35 & 10 & 5 & 6 (2) & window theory; DTFT of $h[n]$ \\
		\bottomrule
	\end{tabular}
\end{table}

\section{Survey Details}
\label{app:itemstats}

\paragraph{Participants.}
\textbf{Students.} We received 28 responses from undergraduate engineering students at EIU. Each student worked through at least one demonstrator, in class or at home, and then completed the questionnaire. The demographic block was optional. Of the 19 students in the quality-screened primary analysis (see Data screening below), 16 filled it in. Of these, six were taking Electromagnetics, three Control Systems, two Engineering Mathematics, one Circuit Analysis, one Signals and Systems, one Digital Signal Processing, and two another course. Seven were in their third year, six in their fourth, and two in their second (one did not answer). Thirteen had opened a single demonstrator and three had opened three or more. Self-reported time on the pages was under 15 minutes for 3 students, 15--30 minutes for 8, 30--60 minutes for 2, and over an hour for 3. Fourteen used a laptop, one a tablet, and one a phone, and fourteen described their MATLAB skill as basic. Time on the pages is self-reported; the opt-in usage logs recorded by the pages were not analysed. The median time to complete the questionnaire was 184\,s. In the full set of 28 responses, 20 completed the demographic block, with the same pattern (Electromagnetics the largest group, eight students; most on a laptop with basic MATLAB skill; median completion time 156\,s).

\textbf{Lecturers.} Twelve lecturers of engineering courses at EIU, who are therefore colleagues of three of the authors, reviewed the six demonstrators, the workflow, and the embedded MATLAB scripts, and then completed a questionnaire that included an independent rubric rating of PPTX, LaTeX/PDF, MATLAB, and HTML. Six had taught for less than five years, two for 5--10 years, and two for 10--20 years (two did not answer). Their courses included Circuit Analysis and Control Systems (five lecturers each), Electromagnetics and Digital Signal Processing (two each), Engineering Mathematics and Signals and Systems (one each), and other subjects (seven). Nine used GenAI at least weekly, four of them daily. Five reported no web-programming skill, five basic skill, and one good skill (one did not answer). Eleven of the twelve already used PPTX in teaching; five used an LMS, four MATLAB, four Jupyter, three LaTeX, and two HTML. The median completion time was 12\,min.

\paragraph{Data screening (post hoc).} After inspecting the responses, we applied one response-quality rule: a student response is excluded if at least 25 of its 26 Likert answers are identical. This rule was not part of the pre-specified analysis plan and is a post-hoc decision. It excludes nine responses: seven answered ``5'' throughout (two of them with a single ``4''), and two answered ``3'' throughout. In both near-invariant responses the differing answer is the first item (SUS1), so a run-based definition (longest run of identical consecutive answers $\ge25$) selects exactly the same nine responses in these data. Invariant patterns are a standard marker of careless responding \citep{meade2012careless}. For the seven responses at ``5'', the pattern is also internally contradictory, because a respondent who strongly agrees both that the page is easy to use (SUS3) and that it is unnecessarily complex (SUS2) is not reporting a usable opinion. Two of the nine took less than 40\,s to answer, and five skipped the demographic block. The remaining 19 responses form the quality-screened primary analysis. As a sensitivity analysis, \cref{tab:sensitivity} repeats the student analysis on all 28 responses, with the same statistics; the substantive interpretation is unchanged, although screening affects the reliability of the SUS markedly ($\alpha=0.11$ with all 28 responses, 0.57 after screening). No lecturer response met the criterion, so all 12 are analysed. Submissions were anonymous and not protected against repeats; the only identical answer vectors in the data belong to the nine excluded responses.

\paragraph{Wilcoxon test settings.} Zero differences are discarded (Wilcoxon's method), and because construct means are heavily tied, $p$-values use the normal approximation with tie correction and no continuity correction (two-sided). The same settings are used for students and lecturers.

\begin{table}[!ht]
	\centering
	\small
	\caption{Sensitivity analysis: student results for all 28 responses, without response-quality screening. Columns as in \cref{tab:survey}.}
	\label{tab:sensitivity}
	\begin{tabular}{lcccccc}
		\toprule
		\textbf{Construct} & \textbf{$\alpha$} & \textbf{Mean} & \textbf{SD} & \textbf{Median} & \textbf{Agree (\%)} & \textbf{$p$} \\
		\midrule
		SI Perceived integration          & 0.89 & 4.17 & 0.76 & 4.38 & 74 & $<0.001$ \\
		SL Perceived learning support     & 0.93 & 4.14 & 0.75 & 4.25 & 75 & $<0.001$ \\
		SE Engagement and interactivity   & 0.92 & 4.19 & 0.73 & 4.33 & 77 & $<0.001$ \\
		ST Trust in simulation            & 0.96 & 4.20 & 0.78 & 4.17 & 76 & $<0.001$ \\
		SP Preference for integrated page & 0.95 & 4.18 & 0.82 & 4.00 & 77 & $<0.001$ \\
		\midrule
		SUS score (0--100)                & 0.11 & 54.5 & 8.1 & 51.3 & \multicolumn{2}{l}{95\% CI 51.3--57.6} \\
		\bottomrule
	\end{tabular}
\end{table}

The invariant responses raise the construct means and the internal consistency of the project scales, but they collapse the SUS reliability, because answering every SUS item identically contradicts its alternating wording. As a further, non-standard check on the screened sample, scoring only the five positively worded SUS items and rescaling them to 0--100 gives 68.2. This figure is not an SUS score and is not comparable with SUS benchmarks; it only shows how much the negatively worded items lower the standard score of 56.8.

\begin{table}[!ht]
	\centering
	\scriptsize
	\setlength{\tabcolsep}{4pt}
	\caption{Item means, standard deviations, and percentage agreeing (4--5) and disagreeing (1--2). Students: $N=19$ after screening. Lecturers: $N=12$. Items marked (R) are negatively worded and shown unreversed.}
	\label{tab:items}
	\begin{tabular}{lcccc@{\hspace{2.2em}}lcccc}
		\toprule
		\textbf{Item} & \textbf{Mean} & \textbf{SD} & \textbf{Agree} & \textbf{Disagree} & \textbf{Item} & \textbf{Mean} & \textbf{SD} & \textbf{Agree} & \textbf{Disagree} \\
		\midrule
		SUS1  & 4.00 & 0.75 & 74 & 0  & LU1 & 4.42 & 0.51 & 100 & 0 \\
		SUS2 (R) & 3.00 & 0.94 & 21 & 32 & LU2 & 4.50 & 0.52 & 100 & 0 \\
		SUS3  & 3.74 & 0.87 & 58 & 5  & LU3 & 4.58 & 0.67 & 92 & 0 \\
		SUS4 (R) & 3.79 & 0.85 & 74 & 11 & LU4 & 4.25 & 0.45 & 100 & 0 \\
		SUS5  & 3.84 & 0.76 & 74 & 5  & LE1 & 3.67 & 0.78 & 67 & 8 \\
		SUS6 (R) & 2.95 & 0.78 & 16 & 26 & LE2 & 3.92 & 0.67 & 75 & 0 \\
		SUS7  & 3.42 & 0.61 & 47 & 5  & LE3 & 3.67 & 1.07 & 67 & 8 \\
		SUS8 (R) & 2.79 & 0.85 & 11 & 32 & LE4 & 3.67 & 0.98 & 50 & 8 \\
		SUS9  & 3.63 & 0.60 & 58 & 0  & LD1 & 3.58 & 0.51 & 58 & 0 \\
		SUS10 (R) & 3.37 & 0.76 & 42 & 11 & LD2 & 4.08 & 0.67 & 83 & 0 \\
		SI1 & 4.16 & 0.69 & 84 & 0 & LD3 & 4.42 & 0.67 & 92 & 0 \\
		SI2 & 4.26 & 0.65 & 89 & 0 & LD4 & 4.00 & 0.60 & 83 & 0 \\
		SI3 & 4.00 & 0.82 & 68 & 0 & LM1 & 3.75 & 0.87 & 67 & 8 \\
		SI4 & 3.53 & 1.02 & 47 & 16 & LM2 & 3.75 & 1.14 & 67 & 8 \\
		SL1 & 4.00 & 0.67 & 79 & 0 & LM3 (R) & 3.75 & 0.87 & 67 & 8 \\
		SL2 & 4.05 & 0.78 & 74 & 0 & LT1 & 3.67 & 0.49 & 67 & 0 \\
		SL3 & 4.00 & 0.75 & 74 & 0 & LT2 & 3.42 & 0.67 & 50 & 8 \\
		SL4 & 3.74 & 0.73 & 68 & 5 & LT3 & 4.33 & 0.78 & 83 & 0 \\
		SE1 & 4.00 & 0.67 & 79 & 0 & LI1 & 3.50 & 0.67 & 42 & 0 \\
		SE2 & 4.11 & 0.74 & 79 & 0 & LI2 & 3.83 & 0.58 & 75 & 0 \\
		SE3 & 3.95 & 0.71 & 74 & 0 & & & & & \\
		ST1 & 3.95 & 0.71 & 74 & 0 & & & & & \\
		ST2 & 4.16 & 0.76 & 79 & 0 & & & & & \\
		ST3 & 4.00 & 0.75 & 74 & 0 & & & & & \\
		SP1 & 3.95 & 0.85 & 74 & 5 & & & & & \\
		SP2 & 4.05 & 0.71 & 79 & 0 & & & & & \\
		\bottomrule
	\end{tabular}
\end{table}

\section{Structure of the Generation Prompts}
\label{app:prompts}

The template below shows the structure shared by the generation requests described in \cref{sec:genai-impl}. It is a description of their form, written for this paper; it is not a verbatim copy of the prompts used.

\begin{enumerate}[leftmargin=1.8em]
	\item \textbf{Context and objectives.} Course, topic, task level (Easy, Medium, or Complex), learning objectives, and the source material (slide text, LaTeX derivation, MATLAB model).
	\item \textbf{Page structure.} A single HTML file with the order objectives, theory, mathematics, visualization, simulation, application, and assessment; the shared plotting, solver, metric, and quiz layer.
	\item \textbf{Mathematics.} Equations in MathJax with the notation of the source, numbered and placed next to the visualization they govern; every displayed number computed from a displayed equation.
	\item \textbf{Visualization and simulation.} The controls and their ranges, the numerical method (e.g., RK4, FFT, Yee update), and what is plotted.
	\item \textbf{Verification.} The reference (analytical or MATLAB), the metric ($E_{\mathrm{rel}}$ or NRMSE), the tolerance, and a PASS/FAIL verdict visible on the page; a collapsible MATLAB script reproducing the reference.
	\item \textbf{Revision.} When a check or review fails: the failing quantity or passage, the expected behaviour, and the instruction to change only the affected part.
\end{enumerate}


\clearpage
\section{Demonstrator Screenshots}
\label{app:screens}

\begin{figure}[!ht]
	\centering
	\begin{subfigure}{0.92\textwidth}\centering
		\includegraphics[width=\textwidth]{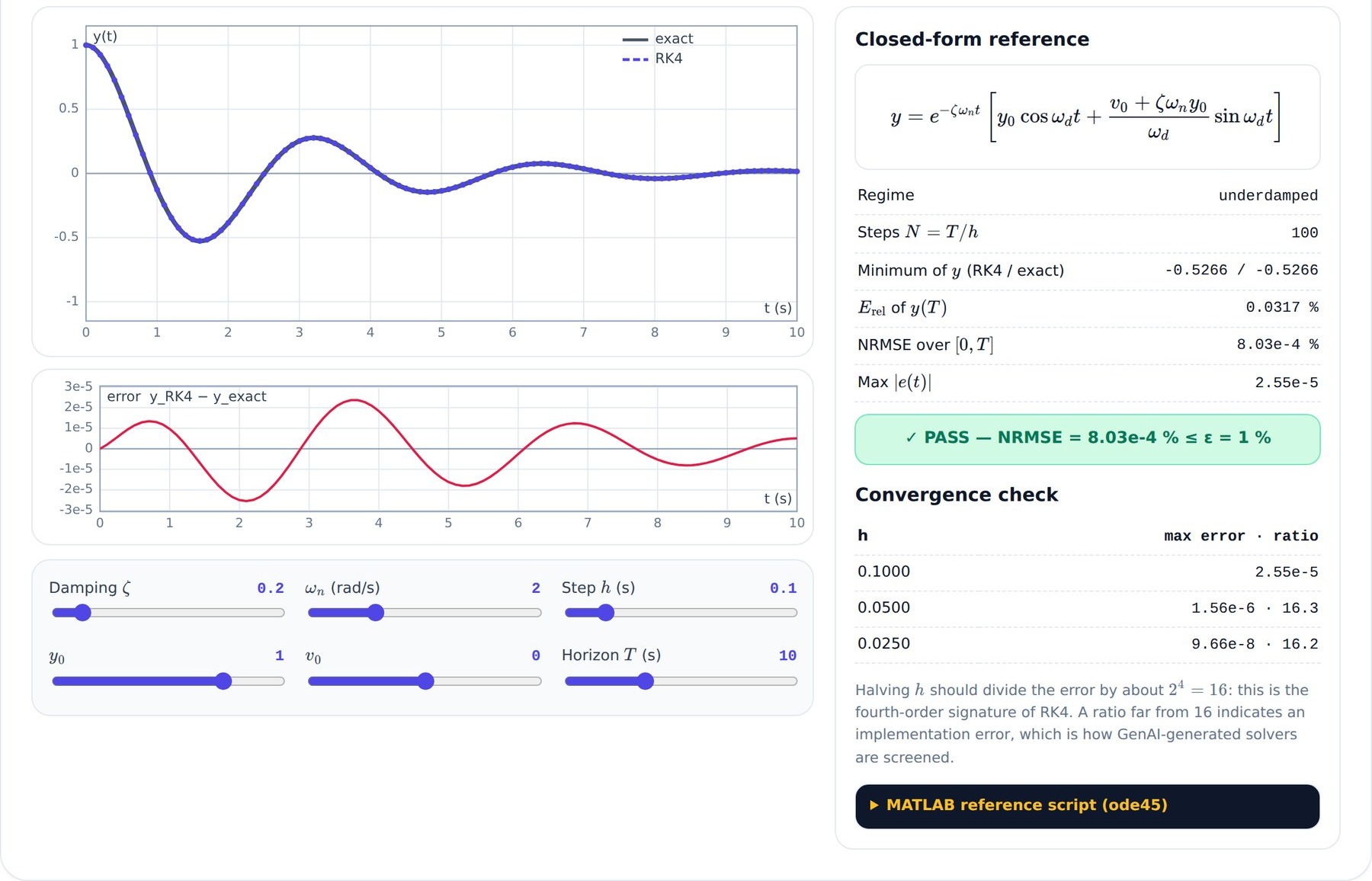}
		\caption{MAT-C: RK4 vs.\ closed form, error plot, verification verdict, and convergence ratio $\approx16$.}
	\end{subfigure}\\[6pt]
	\begin{subfigure}{0.92\textwidth}\centering
		\includegraphics[width=\textwidth]{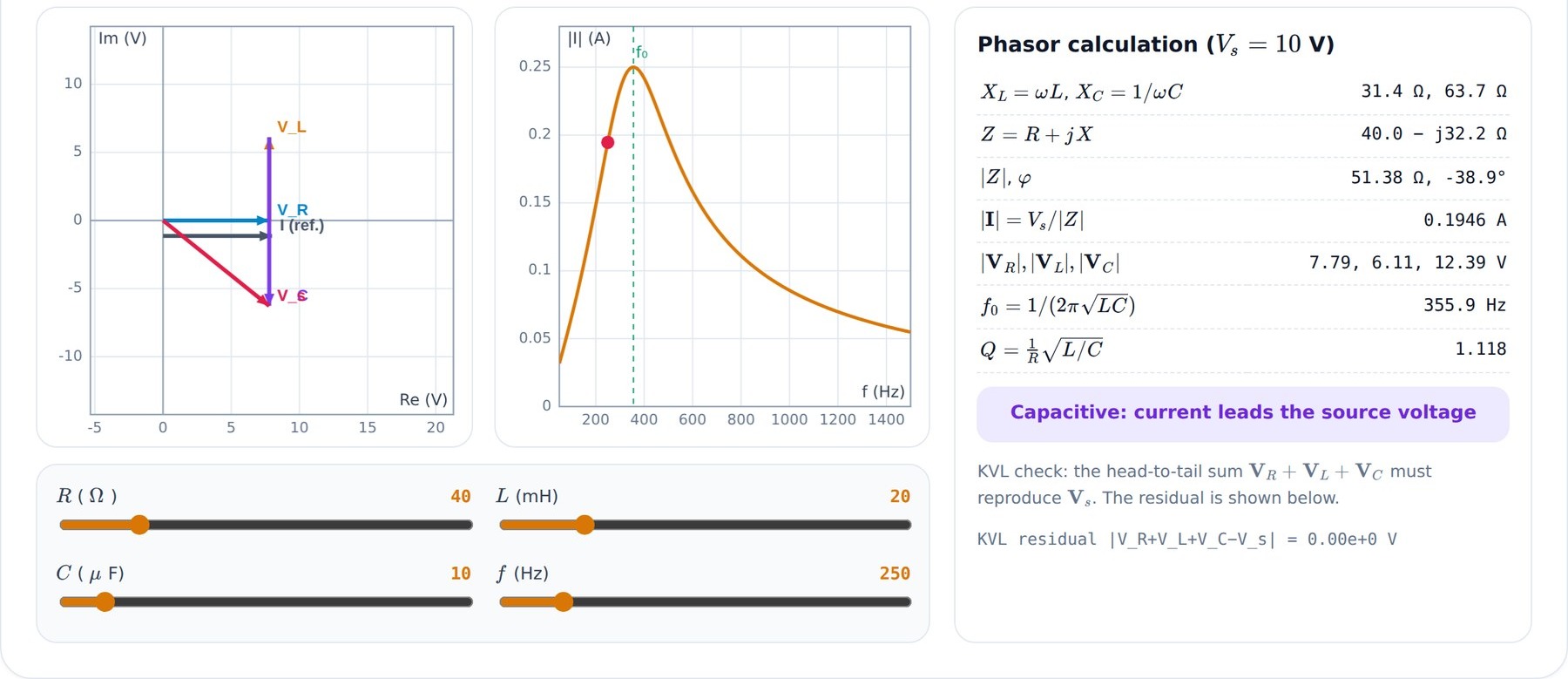}
		\caption{CIR-M: phasor diagram with KVL residual and resonance curve.}
	\end{subfigure}
	\caption{Engineering Mathematics and Circuit Analysis demonstrators. Screenshots rendered in headless Chromium and cropped to the interactive and verification area.}
	\label{fig:demos}
\end{figure}

\begin{figure}[!ht]
	\centering
	\begin{subfigure}{0.92\textwidth}\centering
		\includegraphics[width=\textwidth]{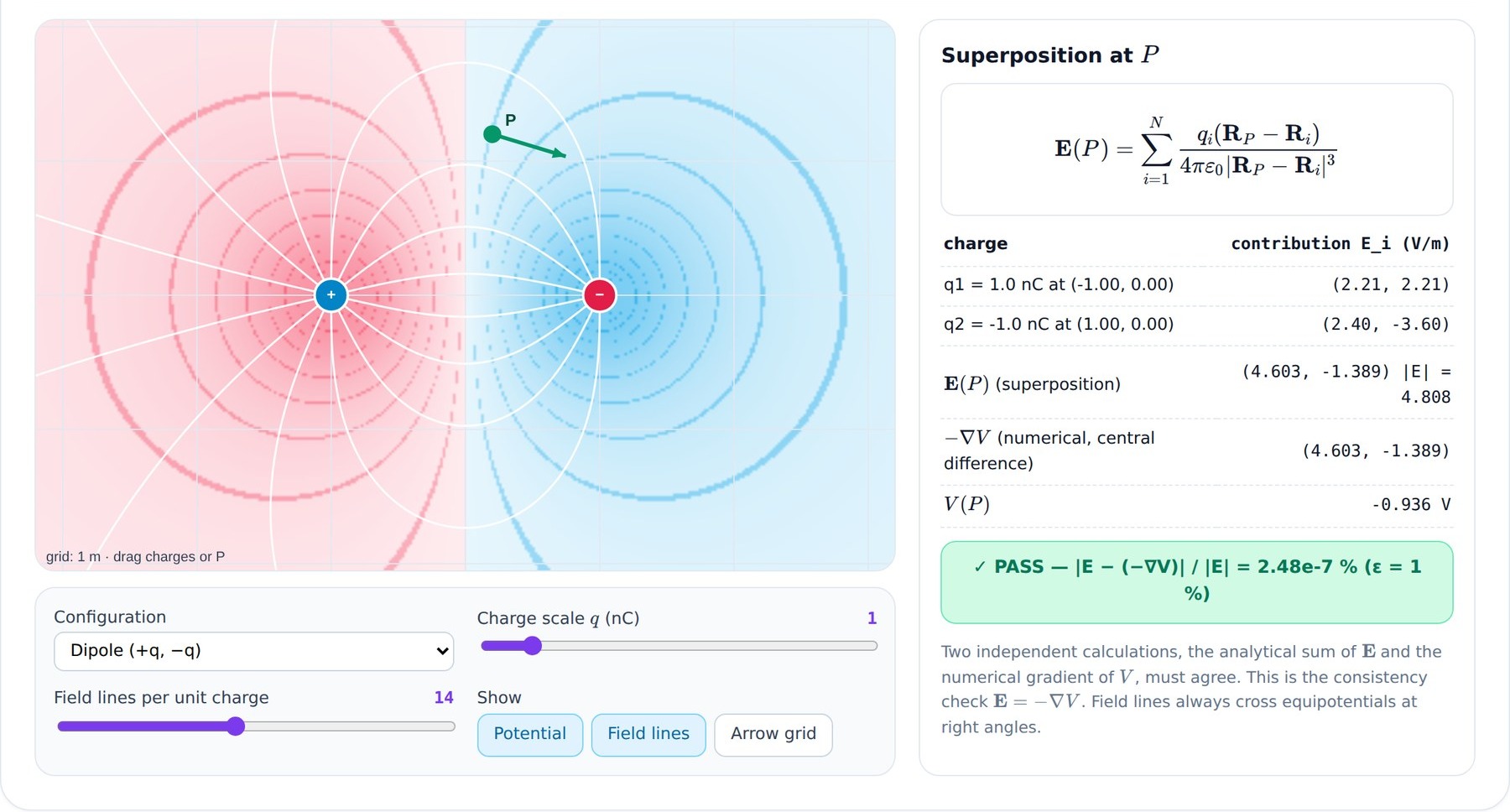}
		\caption{EMF-M: potential map, field lines, and superposition at a draggable probe, checked against $-\nabla V$.}
	\end{subfigure}\\[6pt]
	\begin{subfigure}{0.92\textwidth}\centering
		\includegraphics[width=\textwidth]{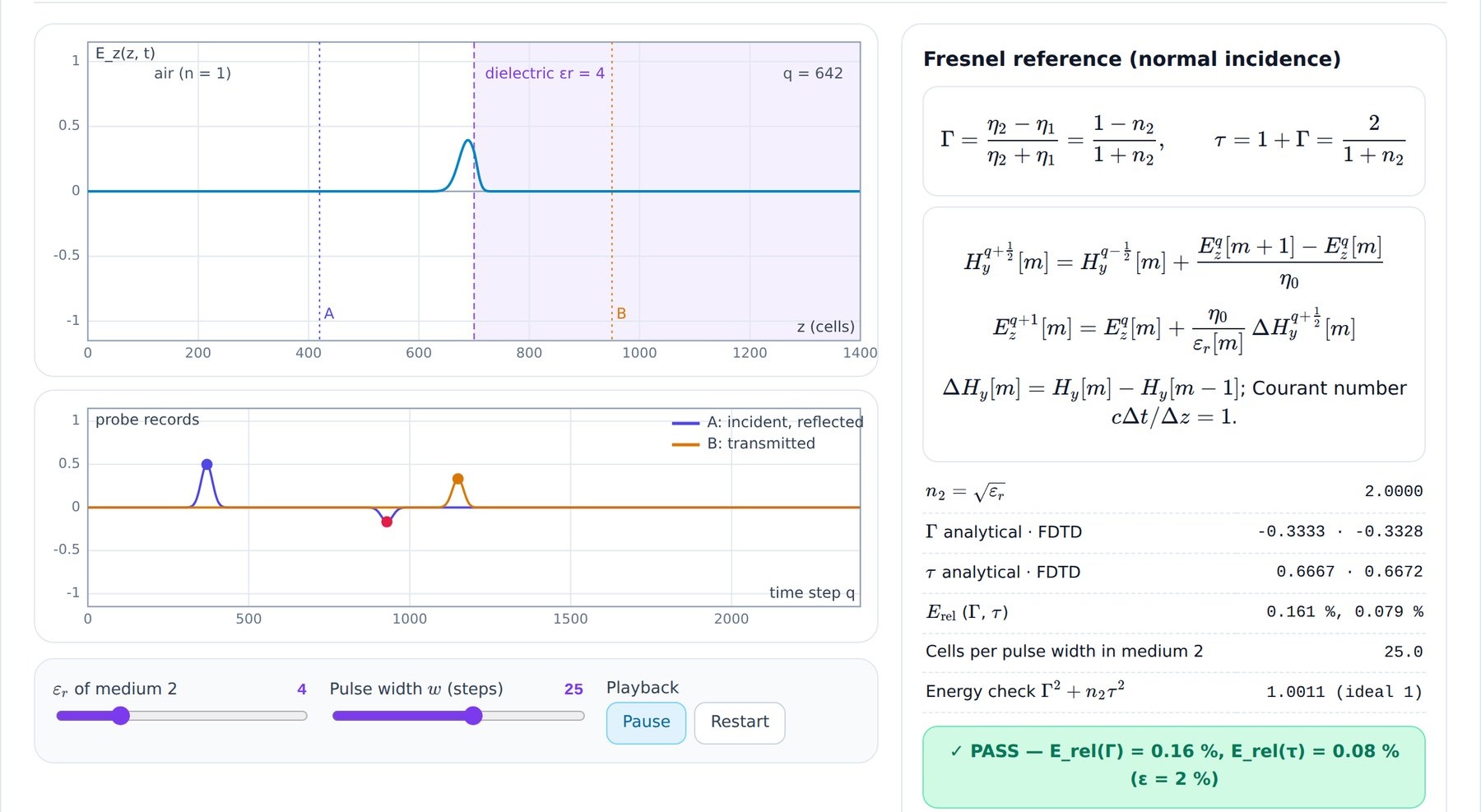}
		\caption{EMF-C: 1D FDTD pulse at a dielectric interface with measured $\Gamma$ and $\tau$.}
	\end{subfigure}
	\caption{Electromagnetics demonstrator. Screenshots rendered in headless Chromium and cropped to the interactive and verification area.}
	\label{fig:demos2}
\end{figure}

\begin{figure}[!ht]
	\centering
	\begin{subfigure}{0.92\textwidth}\centering
		\includegraphics[width=\textwidth]{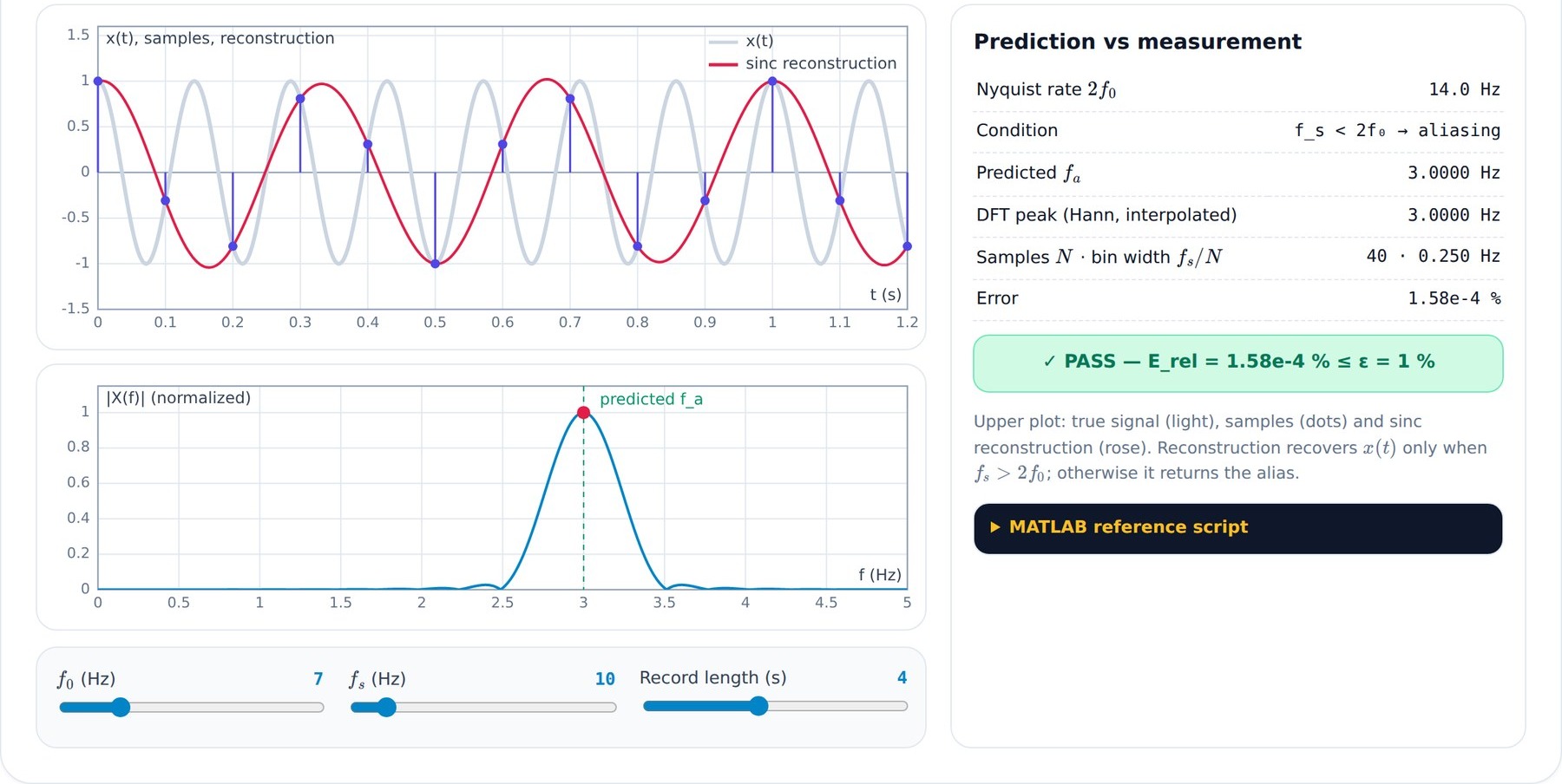}
		\caption{SIG-C: sampling, sinc reconstruction, and DFT-measured alias frequency.}
	\end{subfigure}\\[6pt]
	\begin{subfigure}{0.92\textwidth}\centering
		\includegraphics[width=\textwidth]{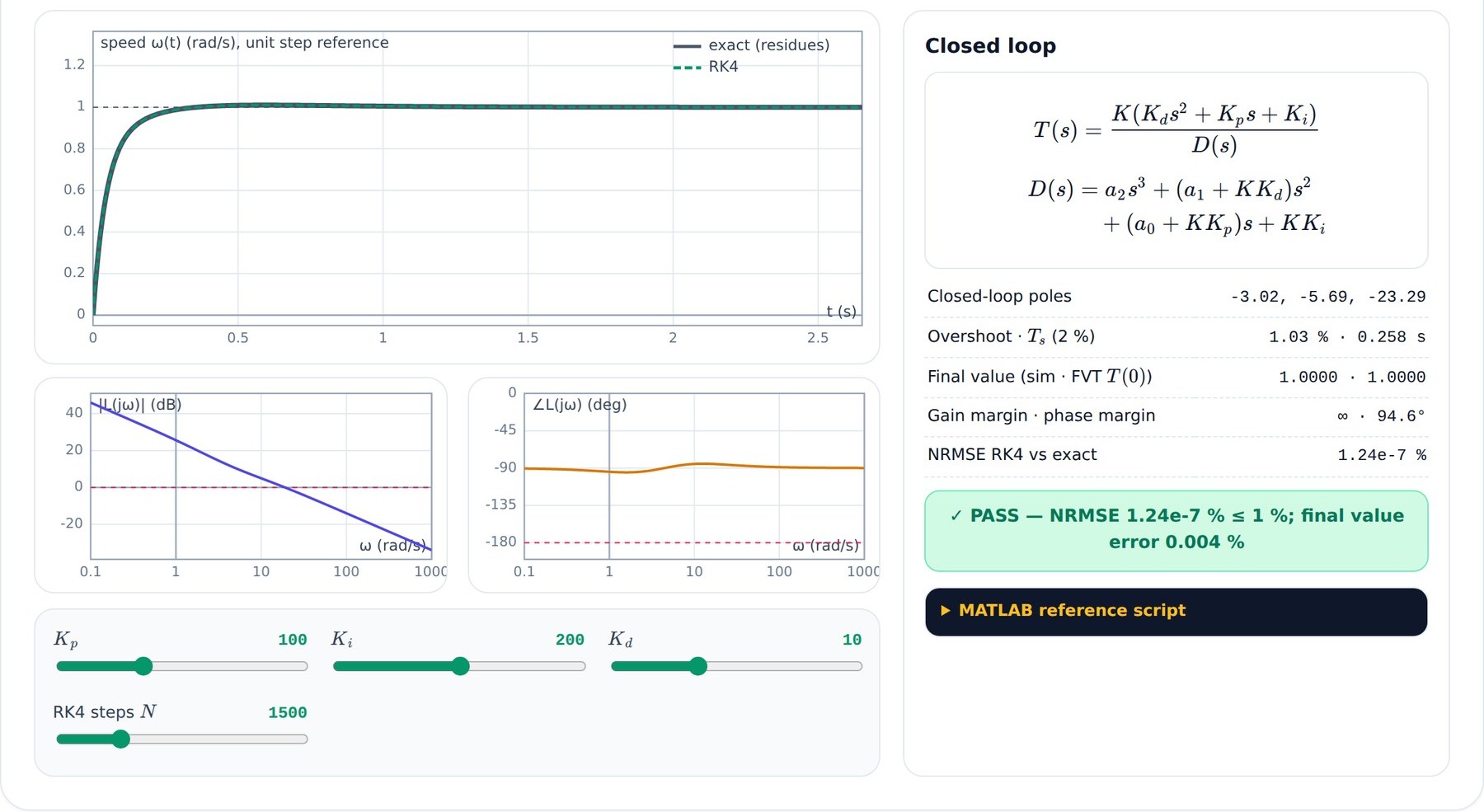}
		\caption{CTL-C: PID speed control of a DC motor with Bode plots and margins.}
	\end{subfigure}
	\caption{Signals and Systems and Control Systems demonstrators. Each Complex section shows its verification verdict on the page. Screenshots rendered in headless Chromium and cropped to the interactive and verification area.}
	\label{fig:demos3}
\end{figure}

\begin{figure}[!ht]
	\centering
	\begin{subfigure}{0.92\textwidth}\centering
		\includegraphics[width=\textwidth]{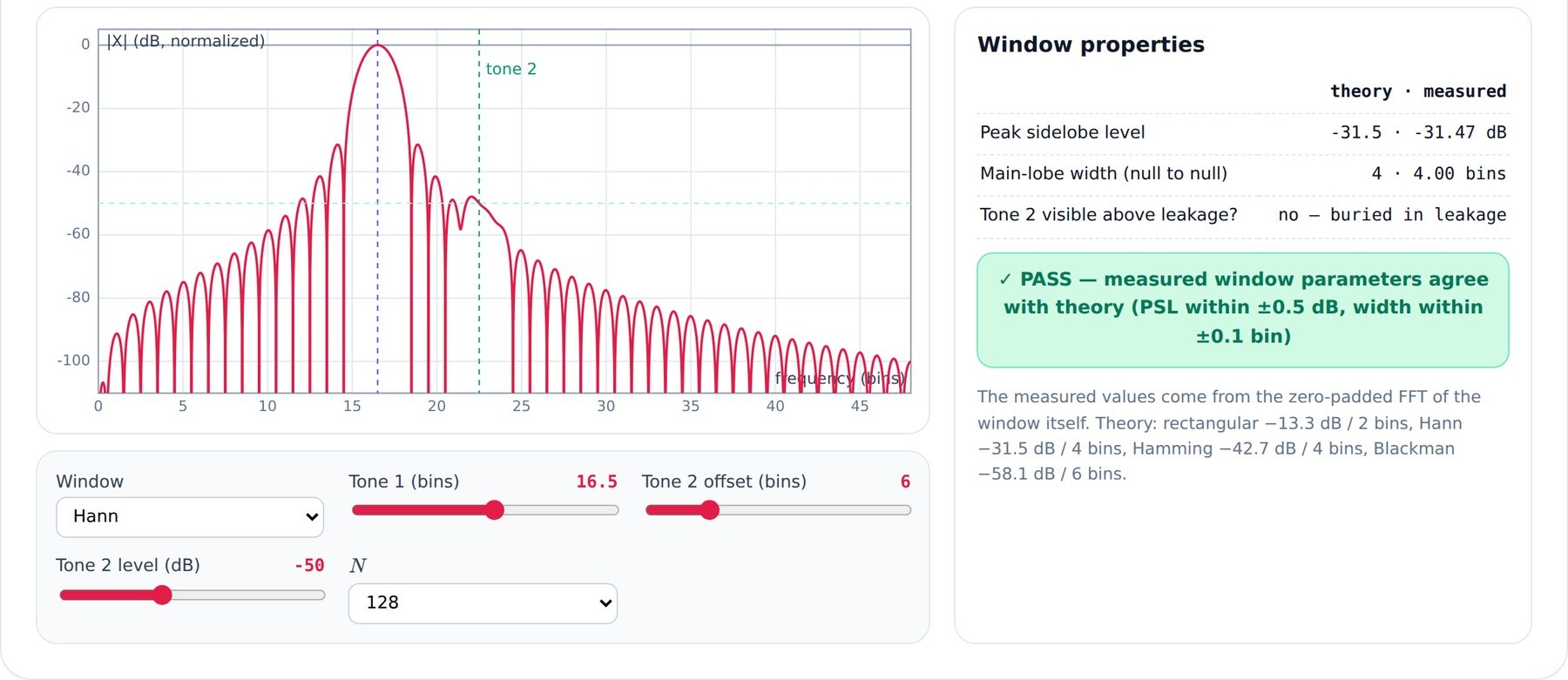}
		\caption{DSP-M: window choice and the visibility of a weak second tone.}
	\end{subfigure}\\[6pt]
	\begin{subfigure}{0.92\textwidth}\centering
		\includegraphics[width=\textwidth]{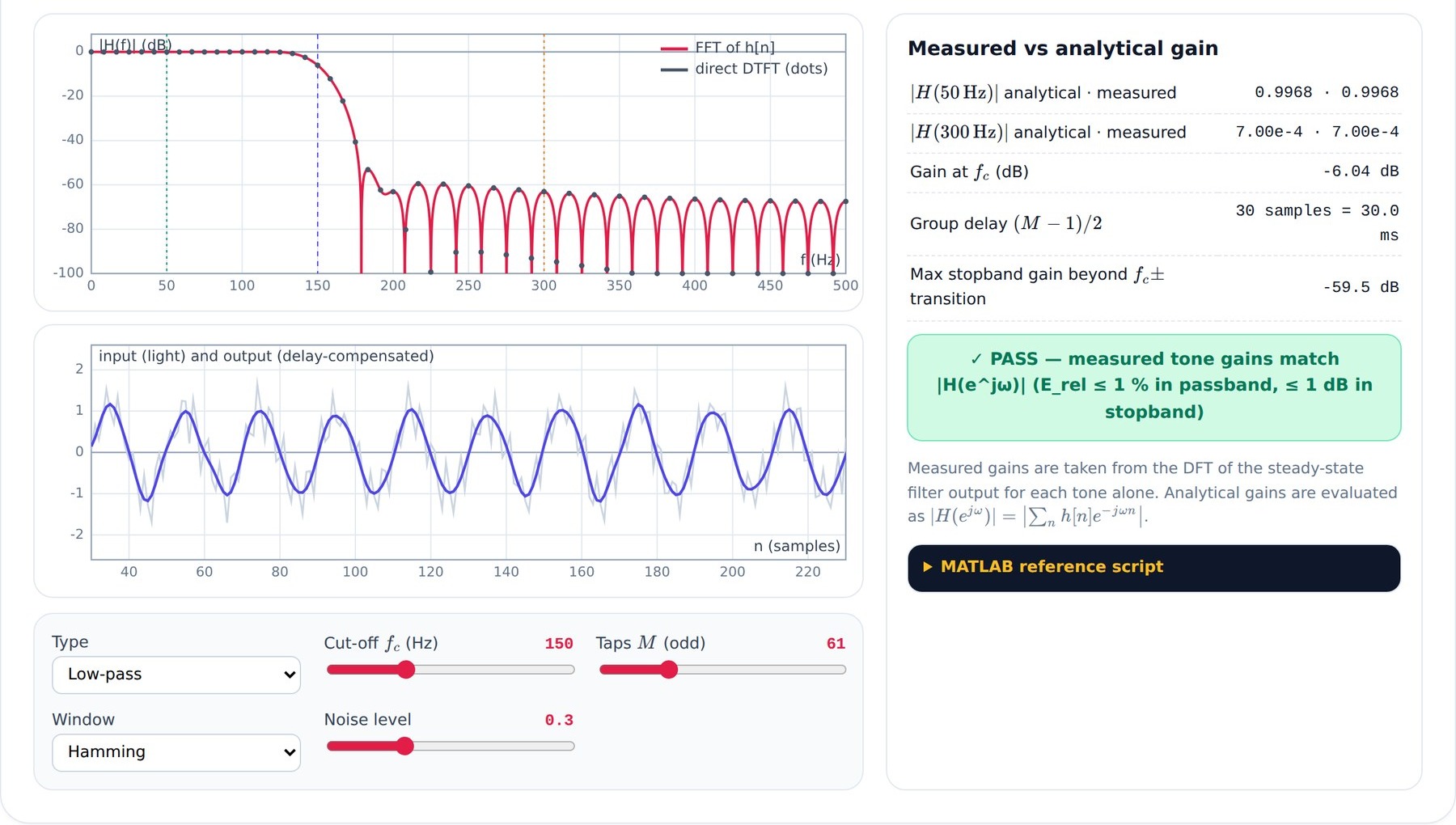}
		\caption{DSP-C: windowed-sinc FIR design, filtering, and gain verification.}
	\end{subfigure}
	\caption{Digital Signal Processing demonstrator. Screenshots rendered in headless Chromium and cropped to the interactive and verification area.}
	\label{fig:demos4}
\end{figure}

Figures \ref{fig:demos}--\ref{fig:demos4} show the interactive and verification areas of the six demonstrators at their default settings.

\end{document}